\documentclass[sigconf,natbib=false]{acmart}
  
\usepackage[utf8]{inputenc}
\usepackage{xcolor}
\usepackage{fancyvrb}
\usepackage{soul}
\usepackage{framed}

\definecolor{summaryback}{gray}{0.95}
\colorlet{shadecolor}{summaryback}

\newenvironment{shadedbox}{%
  \MakeFramed{\advance\hsize-\width\FrameRestore}%
}{%
  \endMakeFramed
}

\usepackage{listings}

\AtBeginDocument{%
  }

\setcopyright{acmlicensed}
\copyrightyear{2026}
\acmYear{2026}
\acmConference[ICSE'26]{IEEE/ACM International Conference on Software Engineering}{April 15--17,
  2026}{Rio de Janeiro, Brazil}

\floatstyle{plain}
\newfloat{listing}{tb}{lst}
\floatname{listing}{Listing}

\newcommand{\hj}[1]{\textcolor{brown}{{\sf HJ}: #1}}

\newcommand{\tool}{\textsc{WhyFlow}{}}
\usepackage{blindtext, graphicx}
\usepackage[labelfont=bf]{caption}
\usepackage{filecontents}
\usepackage{tikz}
\usetikzlibrary{patterns}
\usetikzlibrary{intersections}
\usetikzlibrary{calc}
\usepackage{color}
\usepackage[strings]{underscore}
\usepackage[linesnumbered,ruled]{algorithm2e}
\usepackage{algpseudocode}
\usepackage{url}
\usepackage{footnote}
\makesavenoteenv{table}
\usepackage{amsmath,amsthm}
\usepackage{booktabs}
\usepackage{multirow}
\usepackage{comment}
\usepackage{pifont}
\usetikzlibrary{shadows}
\usepackage{enumitem}
\usepackage{siunitx}
\usepackage{listings}
\usepackage{parcolumns}
\usepackage{multicol}
\usepackage[utf8]{inputenc}
\DeclareUnicodeCharacter{2212}{-}
\usepackage[english]{babel}
\usepackage{colortbl}

\definecolor{dkgreen}{rgb}{0,0.6,0}
\definecolor{gray}{rgb}{0.5,0.5,0.5}
\definecolor{mauve}{rgb}{0.58,0,0.82}
\definecolor{orangep}{rgb}{0.71, 0.43, 0.89}
\definecolor{orp}{rgb}{1, 0.7, 0.278}

\definecolor{darkBlue}{rgb}{0.000000,0.000000,0.545098}
\definecolor{darkGreen}{rgb}{0.000000,0.392157,0.000000}
\definecolor{DarkGray}{gray}{0.4}
\definecolor{javared}{rgb}{0.6,0,0} % for strings
\definecolor{javagreen}{rgb}{0.25,0.5,0.35} % comments
\definecolor{javapurple}{rgb}{0.5,0,0.35} % keywords
\definecolor{javadocblue}{rgb}{0.25,0.35,0.75} % javadoc
\definecolor{lightgray}{gray}{0.95}
\definecolor{shadecolor}{RGB}{150,150,150}
\definecolor{blueA}{RGB}{204,229,255}
\definecolor{redA}{RGB}{112,0, 0}

\algnewcommand\algorithmicforeach{\textbf{for each}}
\algdef{S}[FOR]{ForEach}[1]{\algorithmicforeach\ #1\ \algorithmicdo}

\makeatletter
\newcommand*{\rom}[1]{\expandafter\@slowromancap\romannumeral #1@}

\definecolor{mywhite}{RGB}{255,255,255}
\definecolor{mygray}{RGB}{220,220,220}
\definecolor{olivegreen}{RGB}{0,100,0}

\newif\ifproofread

\proofreadfalse

\begin{document}

\title{\tool{}: Interrogative Debugger for Sensemaking Taint Analysis}

\author{Burak Yetiştiren}
\email{burak@cs.ucla.edu}
\affiliation{%
  \institution{UCLA}
  \city{Los Angeles}
  \state{CA}
  \country{USA}
}

\author{Hong Jin Kang}
\email{hongjin.kang@sydney.edu.au}
\affiliation{%
  \institution{The University of Sydney}
  \city{Sydney}
  \country{Australia}
}

\author{Miryung Kim}
\email{miryung@cs.ucla.edu}
\affiliation{%
  \institution{UCLA}
  \city{Los Angeles}
  \state{CA}
  \country{USA}
}




\begin{teaserfigure}
\centering
  \includegraphics[width=\textwidth]{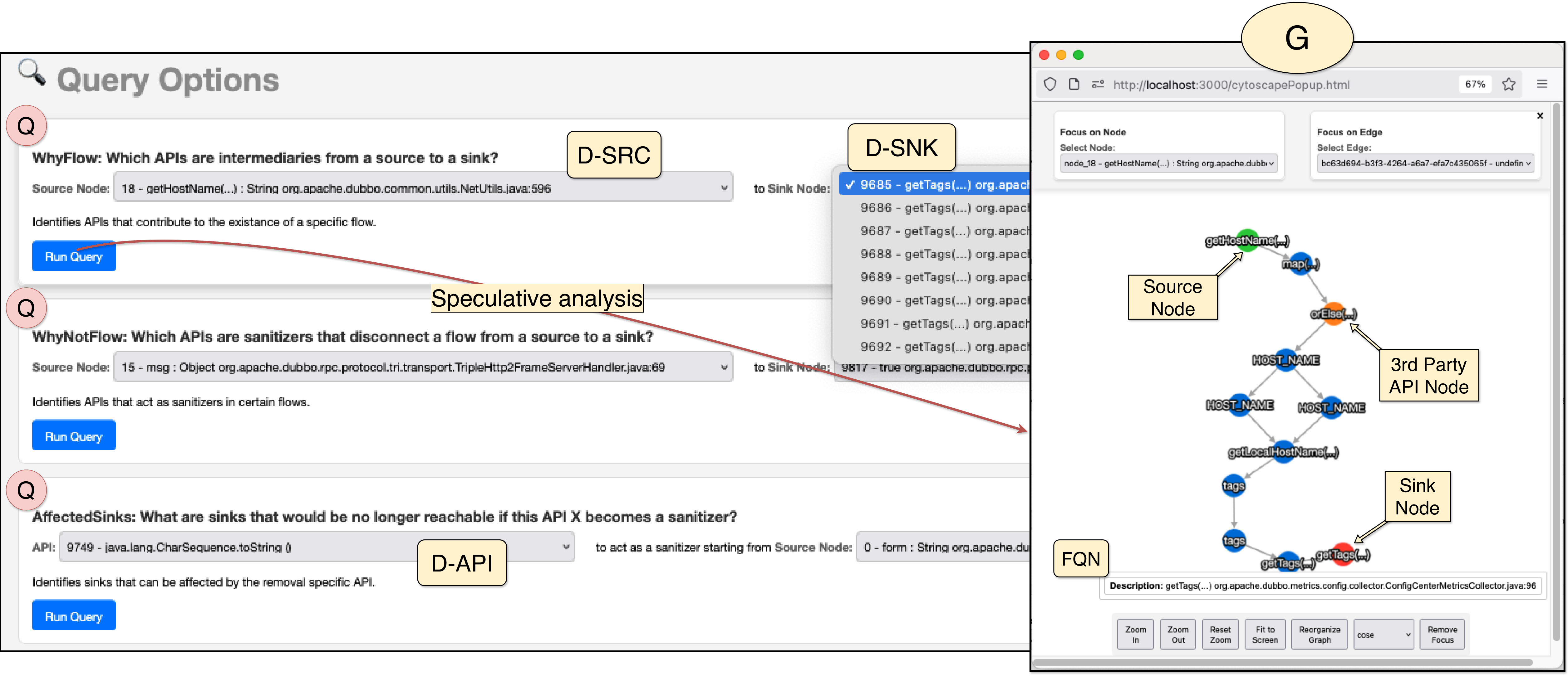}
  \caption{\tool{} supports interrogative debugging by enabling a user to ask \textit{why}, \textit{why-not}, and \textit{what-if} questions about taint analysis. A user can select from templated queries (shown in \texttt{Q}) and contextualize their inquiry with respect to a specific source, sink, and third-party library model using a drop-down menu, shown in (\texttt{D-SRC}), (\texttt{D-SNK}), and (\texttt{D-API}). Once configured, a background question-and-answer analysis is conducted to help a user with their sensemaking process of taint analysis results. To aid in their sensemaking process about global connectivity, permissible and impermissible data flows, a user can see the result in a graph view (\texttt{G}) with color-coded annotation.}   
  \label{fig:teaser}
\end{teaserfigure}

\begin{abstract}
    Taint analysis is a security analysis technique used to track the flow of potentially dangerous data through an application and its dependent libraries. Investigating why certain unexpected flows appear and why expected flows are missing is an important sensemaking process during end-user taint analysis. Existing taint analysis tools often do not provide this end-user debugging capability, where developers can ask \textit{why}, \textit{why-not}, and \textit{what-if}  questions about dataflows and reason about the impact of configuring sources and sinks, and models of third-party libraries that abstract permissible and impermissible data flows. Furthermore, the tree-view or list-view used in existing taint analyzer visualizations makes it difficult to reason about the global impact on connectivity between multiple sources and sinks.

    Inspired by the insight that \textit{sensemaking} tool-generated results can be significantly improved by a QA inquiry process, we propose \tool{}, the first end-user question-answer style debugging interface for taint analysis. It enables a user to ask \textit{why}, \textit{why-not}, and \textit{what-if} questions to investigate the existence of suspicious flows, the non-existence of expected flows, and the global impact of third-party library models. \tool{} performs \textit{speculative what-if} analysis, to help a user in debugging how different connectivity assumptions affect overall results. A user study with 12 participants shows that participants using \tool{} achieved 21\% higher accuracy on average, compared to CodeQL. They also reported a 45\% reduction in mental demand (NASA-TLX) and rated higher confidence in identifying relevant flows using \tool{}. 
% Most critically, participants using \tool{} were 31\% more accurate in answering questions about how the models of external libraries impact the reported flows. 
This shows \tool{}’s potential to significantly reduce sensemaking effort.   
\end{abstract}

\maketitle

\section{Introduction}\label{sec:introduction}
% 1: static analysis, taint analysis. 

To prevent security vulnerabilities, developers use taint analysis. This technique tracks potentially dangerous data as it moves through a program. Data from untrusted sources, like user input, is marked as ``tainted.'' The analysis then monitors how this tainted data spreads. If it reaches a critical point in the code, known as a ``taint sink,'' a warning is generated. Tainted data can be made safe by ``sanitizers,'' which are functions that clean or encode the data. For example, the \hl{\texttt{java.net.URLEncoder.encode(String input, String encoding)}} function escapes potentially harmful characters, preventing security issues.
Since analyzing the entire program is usually impossible, practical taint analysis requires configuring simplified models of external libraries~\cite{Chibotaru2019}. 
These models abstract the behavior of third-party libraries, allowing an analysis to track taint flow without analyzing the libraries' internals.

% 2. how modeling affects the result
Taint analysis tools often come with default configurations, including pre-built {\em models of external libraries}. However, these default models, which are often automatically generated, can be inaccurate. 
This is because they rely on assumptions, which may be incorrect, about how libraries handle data flow.
% , and these assumptions may not always hold true. 
% The taint analysis result is highly dependent on these modeling assumptions and configurations. 
Incorrect configurations can lead to significant problems. 
For example, if a model incorrectly allows tainted data to flow through a library function that should sanitize it, the analysis will produce unexpected taint flows (i.e., false positives)~\cite{Livshits2009, Banerjee2023}.
This means it will report potential vulnerabilities that do not actually exist, creating extra, unexpected warnings.
% for the user. 
Conversely, if a model incorrectly blocks the flow of tainted data when it should not, the analysis will produce missing flows (i.e., false negatives). 
This means it will miss real vulnerabilities, failing to report flows expected by the user.

We introduce \tool{}, a novel end-user debugger that brings interactive, question-based sensemaking to taint analysis. 
Drawing inspiration from interrogative debugging principles, \tool{} enables developers to ask \textit{why}, \textit{why not}, and \textit{what if} questions about their analysis configuration.
This helps developers make sense of the impact of configuration and models of third-party libraries on tool-generated warnings. 
This sensemaking process involves tracing the reported dataflows, understanding why specific warnings are generated, and how these warnings relate to the models of third-party libraries.
\tool{} focuses on debugging an end-user configuration instead of a taint analysis implementation.

\tool{} provides six customizable question templates, which are then concretized by a user. A user can make sense of currently detected taint flows and hypothetical flows through interactive QA. Unlike state-of-the-art tools like CodeQL that rely on list views, \tool{} visualizes these reported and hypothetical flows graphically, offering a more intuitive understanding of configuration changes. Below, we discuss two key features: \emph{Inquiry-based Sensemaking} and \emph{Visualization of Global Impact} in detail. 

\textbf{Inquiry-based Sensemaking.} To effectively debug taint analysis results, developers need to investigate why specific dataflows are reported and why others are not. Unfortunately, many tools do not allow developers to interactively ask questions about these flows. Furthermore, to efficiently narrow down their analysis, developers must understand how modifying the models of external libraries impacts the permitted or blocked dataflows.

In the \textit{Query Options} pane of \tool{} (shown in Figure~\ref{fig:teaser}), users can select from pre-defined question templates. These templates allow users to investigate taint analysis results by asking specific questions about dataflows. To make these questions concrete, users specify the configuration they are interested in, including the source of the data, the destination (sink), and the models of third-party libraries involved.

For example, a user might choose the \textit{why-flow} template, which asks, ``Why is there a taint flow from a source to a sink?'' They can then concretize this question by specifying a specific source and sink.
For instance, they could ask, ``Which third-party library models currently allow taint flows from the source \texttt{java\-.net\-.Inet\-Address\-.get\-Host\-Name(...)} to the sink \texttt{org\-.apache\-.dubbo\-.\-met\-rics\-.model\-.Config\-Center\-Metric\-.\-getTags\-(\ldots)}?''

Similarly, a user could select the \textit{why-not} template, which asks, ``Why is there no taint flow from a source to a sink?'' They can then specify the source and sink of interest.
For example, they could ask, ``Which third-party library models could potentially block taint flows from the source \texttt{msg\-:Http\-Request} to the sink \texttt{Error\-Type\-Aware\-Logger\-.warn()}?'' 
Alternatively, a user could also select the \textit{what-if} template, which asks a speculative analysis question: ``Which sinks would no longer be reachable if third-party library models were configured as a sanitizer from source to sink?''

%\begin{figure*}
%    \centering
%    \includegraphics[width=\textwidth]{images/query_options.png}
%    \caption{\textit{First two query options available in \tool{}.} Here the user can select the configuration (source, sink, 3\textsuperscript{rd}-party library) options from the related drop-down menu. After the selection, the query can be initiated by clicking the ``Run Query'' button for the given question about the selected taint flow.}
%    \label{fig:query-options}
%\end{figure*}

\textbf{Visualization of Global Impact.} Once a user initiates a query within \tool{}, specifying their target source, sink, and any relevant external API calls, \tool{} generates an interactive graph visualization, providing a holistic view of the taint flow. This graph, as seen in the \texttt{G} pane of Figure \ref{fig:teaser}, visually maps the program's taint flows as interconnected nodes, each representing a distinct stage in the taint flow.
This visualization reveals the {\em global impact of configuration choices} through the network of connected nodes.
% To facilitate sensemaking, 
To draw attention to the impact of configuration choices,
\tool{} employs a color-coded scheme: source nodes are highlighted in green, sink nodes in red, and external library nodes in orange.
% \sout{.
% This color differentiation draws attention to the impact of configuration choices}.
% \sout{Furthermore, t}
Templated queries are designed to illuminate how third-party library models can impact multiple sinks and multiple sources, making the global impact salient through visualization.

\textbf{User study:} We conducted a within-subject study with a factorial crossover involving 12 participants (graduate students and professional developers) who inspected taint-analysis warnings generated by CodeQL~\cite{avgustinov2016ql,szabo2023incrementalizing}.
Each participant answered eight questions per task (16 total) designed to reflect a realistic sensemaking process of tool-generated warnings
% During
% these
% \textcolor{red}{each of the \sout{this}}
% tasks, a user 
They
had to answer questions about the impact of third-party library models on taint analysis results by identifying pass-through APIs calls, APIs that serve as sanitizers, etc. Participants increased the average question completion rate from 71\% with the baseline CodeQL visualizer to 92\% with \tool{}, especially on the kinds of questions that require global reasoning of multiple taint flows.
% originating from the same source to multiple sinks, or multiple taint flows reaching the same sink starting from different sources. 
When using \tool{}, users were more accurate in identifying multiple sinks affected by the same third-party library model
% \sout{.
% For example, 50\% of answers prepared with \tool{} are correct, whereas only 17\% with the baseline} 
(50\% vs.\ 17\%).

We measured cognitive load via NASA-TLX questions. We found that \tool{} led to improvement on the participants' self-reported mental demand, stress, and success rate. 
Considering the technology acceptance model~\cite{lee2003technology}, participants rated \tool{} higher on both \emph{Confidence} (mean 4.3 vs.\ 2.1) and \emph{Ease of Use} (4.3 vs.\ 1.9) than the baseline CodeQL. 
Qualitative analysis highlighted that color-coded graphs and template queries helped alleviate {\em analysis paralysis}~\cite{schwartz2015paradox}. 
% by grouping related warnings and clarifying how multiple taint-flows are affected by the configuration of third-party models, sources, and sinks. 
Overall, these results underscore that \tool{} substantially enhances users’ ability to make sense of complex taint paths and reduce their mental workload through QA-based debugging.
% through QA-based end-user debugging. 

In this paper, we make the following contributions:
\begin{enumerate}
    \item Configuring taint analysis is tricky, often leading to unexpected results. 
    To help users understand and fix these issues, \tool{} is the first interactive end-user debugger that shows how different user configuration choices and modeling of third-party libraries impact the taint analysis results.
    \item \tool{} is equipped with template queries for {\em `why,'} {\em `why not,'} and {\em `what if'} questions, which are automatically translated into logic queries; it enables users to concretize the template queries with concrete code names embedded in tool-generated warnings. \tool{} makes it easier to examine the global impact of models, which are difficult to reason when viewing warnings in a list view, packaged with an existing taint analyzer. 
    \item We conduct a within-subject user study with a factorial crossover design (12 participants) to assess \tool{}'s effectiveness in sensemaking CodeQL-generated taint analysis warnings~\cite{avgustinov2016ql,szabo2023incrementalizing}. \tool{} raises the average question completion rate from 71\% to 92\% and reduces mental demand (NASA-TLX) from 5.9 to 3.3, demonstrating clear improvement in the end-user debugging of taint analysis.
    \item Improvement in accuracy, reduction in cognitive load, improvement in confidence, and ease of use compared to CodeQL's visualizer is statistically significant. The additional follow-up study with two professionals corroborates this improvement in accuracy, confidence, and ease of use. 
\end{enumerate}

The remainder of this paper is organized as follows.
Section~\ref{sec:motivation} introduces a motivating example and the design goals. Section \ref{sec:approach} presents our tool, \tool{}.
Section~\ref{sec:userstudy} provides the study design. Section~\ref{sec:results} reports the user study's results.
Section~\ref{sec:discussion} discusses the implications of our findings and threats to validity. Section~\ref{sec:relatedwork} presents related work.
Section~\ref{sec:conclusion} concludes our paper.
% \by{Add a screenshot of CodeQL§}\by{Done.}

\section{Motivation}\label{sec:motivation}
% 1.  lack of end-user debugging support. 
When using modern static taint analyzers such as CodeQL~\cite{szabo2023incrementalizing},  
developers struggle to identify why unexpected flows appear or why certain expected flows are missing.
These tools rely on models of external libraries, which may not always match the developer's expectations. 
This paper is concerned with this problem of {\em end-user debugging of taint analysis}\textemdash i.e., understanding the impact of user configuration choices. 
% as opposed to debugging analyzer implementation.      
Two key end-user debugging challenges prevent developers from easily making sense of unexpected results.

First, traditional tools lack support for inquiry-based debugging, presenting only isolated warnings without contextual flow connections. This forces developers into manual tracing, which
% This is tedious and
hinders their ability to answer questions they have during debugging.
% \sout{.
% A robust QA process, focused on sensemaking, is therefore essential to bridge this gap and enable active investigation rather than passively viewing the flow}. 
% , enabling developers to actively investigate and understand the underlying logic rather than passively viewing dataflow visualizations. 

Second, beyond isolated flow views,
developers need support for reasoning about the model's impact on multiple dataflows. Current tools lack the ability to predict the ripple effects of model changes,
forcing developers to manually piece together disparate flows.
% , hindering them from understanding the consequences of model changes. 
% They need a debugger that enables comprehensive, connectivity-driven reasoning to truly grasp the global implications of third-party library models.

% forced to laboriously re-run analyses (often with only minimal hints from the tool) to assess whether altering certain assumptions addresses or corrects a purported security flaw.
%\miryung{instead of static analysis's viewing interface format, we should argue that existing tools do not allow interrogative debugging capability. e..g cannot easily ask questions about how different assumptions about 3rd party libraries' models and different assumptions about reachability  affect the overall findings}\by{Updated.}

% \begin{enumerate}
%     \item \textbf{Insufficient QA capability about alternative data flow assumptions.} 
%     \miryung{I likt}
%     \item \textbf{Inaccurate Models of Third-Party Libraries.} External libraries are frequently over- or under-approximated, leading to false positives and unreported vulnerabilities. 
% \end{enumerate}

Prior work introduced \emph{WhyLine}~\cite{ko2004designing} to let users ask \emph{``why did''} or \emph{``why didn't''} questions on a program trace, for debugging runtime errors; however, \emph{``why and why-not''} questions remain unaddressed in the context of modern taint analysis. Developers want customizability over taint analysis~\cite{nachtigall2022large}, visual outputs of warnings~\cite{johnson2013don}, and the ability to inspect intermediate pass-through nodes of the analysis~\cite{ko2004designing}. Yet, typical dataflow or taint analysis interfaces do not enable higher-level inquiries such as ``\emph{Why does data from source~X reach sink~Y?}'' or ``\emph{Why is no warning raised for a known bug?}'' 

%The \textit{WhyFlow} query identifies the APIs that act as intermediaries, and permit the data flow between a given source $X$ and sink $Y$. 
%In contrast, the \textit{WhyNotFlow} query reveals which 3\textsuperscript{rd}-party API call is acting as a sanitizer that is disrupting or disconnecting the data flow between the selected source and sink. 
%To understand the impact of potential changes, the \textit{AffectedSinks} query determines which sinks would become unreachable if a specific 3\textsuperscript{rd}-party API $Z$ were to serve as a sanitizer, thereby terminating the data flow from source $X$. 
%Furthermore, the \textit{DivergentSinks} query highlights the common intermediaries shared between a source and a pair of sinks, offering insights into convergent flow patterns. Similarly, the \textit{DivergentSources} query uncovers which intermediaries are common to a pair of sources converging on a single sink. 
%Finally, the \textit{GlobalImpact} query ranks intermediary 3\textsuperscript{rd}-party APIs based on their frequency of occurrence in data flows from $X$ to $Y$, providing a broader perspective on their overall impact. 
%Together, these queries offer a robust framework for pinpointing the impact of third-party library models in complex taint flows. 

\begin{table*}[t]
  \caption{Template queries for {\it why}, {\it why-not} and {\it what-if} questions and corresponding English interpretation and logic queries.}
  \label{tab:queries}
  \centering
  \begin{tabular}{lp{0.35\textwidth}p{0.45\textwidth}}
    \toprule
    \textbf{Query Type} & \textbf{Plain-English Question} & \textbf{Logic Query Interpretation}\\
    \midrule
    \textsc{WhyFlow}
    & ``Why is there a taint flow from a source $X$ to a sink $Y$?''
    & ``Which third-party library models (or assumptions) currently allow data to propagate from $X$ to $Y$?'' \\[2pt]

    \textsc{WhyNotFlow}
    & ``Why is there \emph{no} taint flow from a source $X$ to a sink $Y$?''
    & ``Which third-party library models (or assumptions) currently terminate the flow (e.g., sanitizers), where their model change could create a flow from a source $X$ to a sink $Y$?'' \\[2pt]

    \textsc{AffectedSinks}
    & ``If we alter a third-party library $Z$'s model, which sinks are affected?''
    & ``Under a new assumption of treating a third-party library $Z$ as a sanitizer, which previously reported sinks are no longer reachable from $X$?'' \\[2pt]

    \textsc{DivergentSinks}
    & ``Which third-party library $X$'s model could influence  multiple taint flows from the same source $X$?'' 
    & ``What are the common third-party API nodes in multiple paths originating from $X$ that \emph{split} into multiple different sinks?'' \\[2pt]

    \textsc{DivergentSources}
    & ``Which third-party library $X$'s model could influence  multiple taint flows reaching the same sink $Y$?'' 
    & ``What are the common third-party API nodes in multiple flow paths that eventually reach the same sink $Y$?'' \\[2pt]

    \textsc{GlobalImpact}
    & ``Which third-party library $Z$'s model could have the largest global influence on dataflows from $X$ to $Y$?''
    & ``What is the frequency of each third-party API call appearing along all paths from $X$ to $Y$ in terms of the overall usage counts?'' \\
    \bottomrule
  \end{tabular}
\end{table*}

% \emph{TraceLens} steps beyond mere visualization to serve as a \emph{question-and-answer (Q\&A) tool} for interactive, inquiry-based debugging of taint flows. 

Table~\ref{tab:queries} shows six kinds of templated queries. In the following paragraphs, we detail the motivating scenario behind each query. 

\begin{figure}
    \centering
    \includegraphics[width=\linewidth]{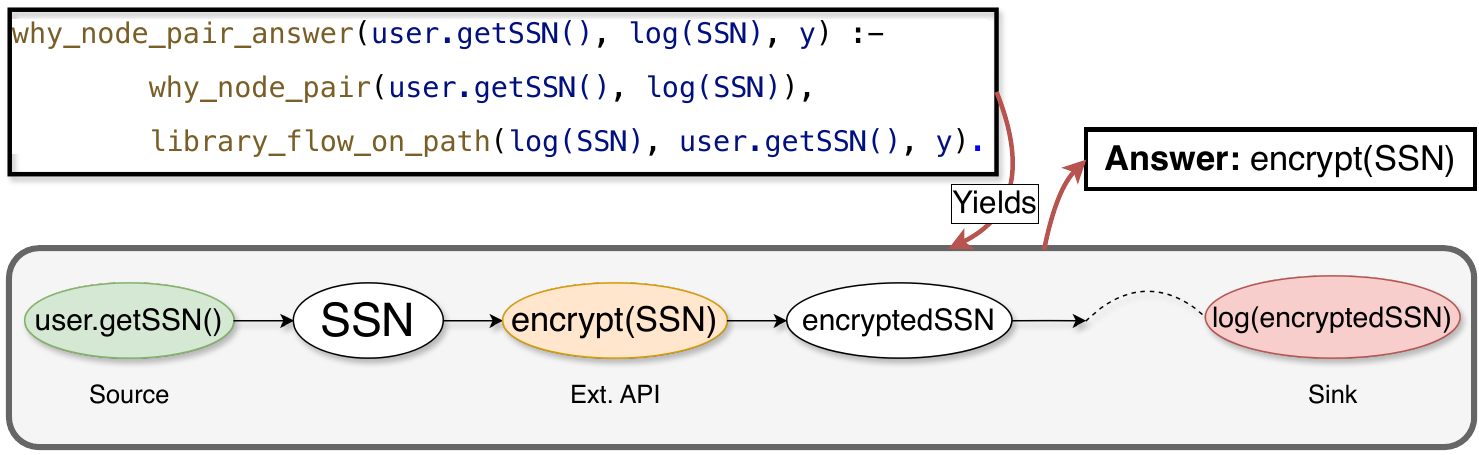}
    \caption{\textsc{WhyFlow}: ``Why is there a taint flow from a source to a sink?''}
    \label{fig:query-whyflow}
\end{figure}

\noindent{\textbf {1. \textsc{WhyFlow}: Why is there a taint flow from a source X to a sink Y?}}
Suppose that Alice sees an unexpected taint-analysis warning in CodeQL from a source \texttt{user.getSSN()} (untrusted data) to a sink \texttt{log(SSN)} (potential vulnerability)~\cite{ko2004designing,nachtigall2022large,johnson2013don}. She suspects the culprit is an imprecise third-party library model; however, manually tracing a long chain of calls is overwhelming. \textsc{WhyFlow} query highlights the taint flow path from a source in \textcolor{green}{green} to a sink in \textcolor{red}{red}, displaying pass-through third-party API calls in \textcolor{orange}{orange}, shown in Figure~\ref{fig:query-whyflow}. 
Alice notices an intermediate API call \texttt{encrypt(SSN)}.
%Alice then notices an intermediate API call is \texttt{encrypt(SSN)}. 
She then discovers that this API is a sanitizer, and thus this warning should \emph{not} have been reported and \texttt{encrypt()}'s model should be debugged.

\begin{figure}
    \centering
    \includegraphics[width=\linewidth]{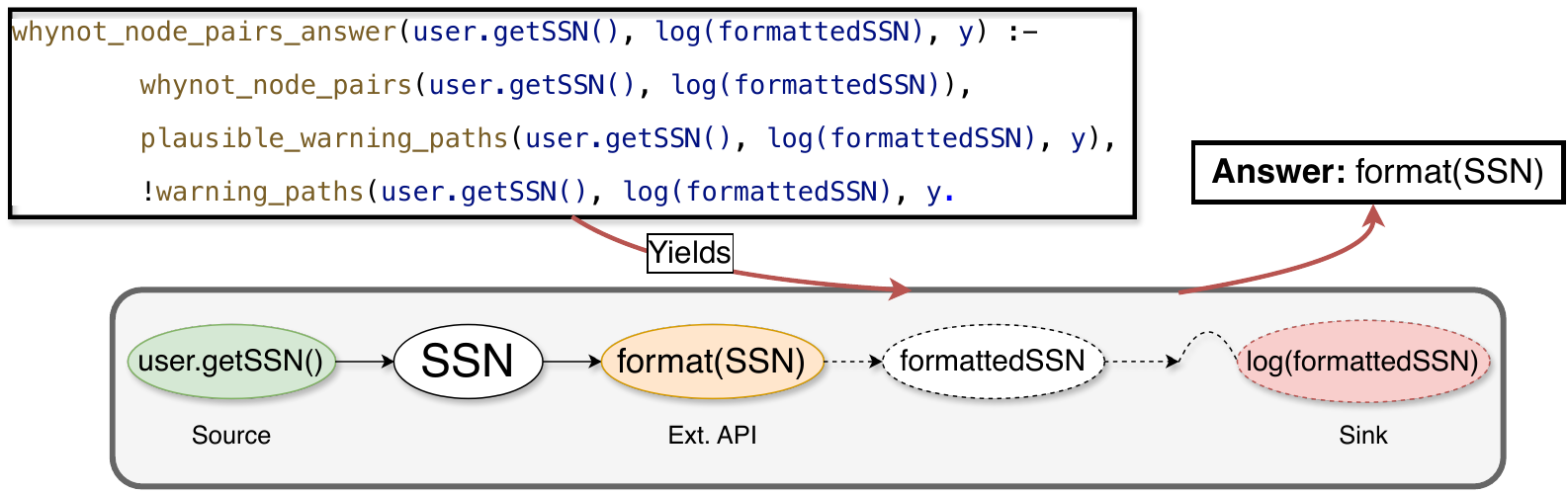}
    \caption{\textsc{WhyNotFlow}: ``Why is there no taint flow from a source to a sink?''}
    \label{fig:query-whynotflow}
%    \miryung{Burak, for Figure 3, add a corresponding logic query, add a text box "Answer: format(SSN) is errorneously marked as a sanitizer."}
\end{figure}

\noindent\textbf{2. \textsc{WhyNotFlow}: Why is there no taint flow from a source X to a sink Y?}
Suppose that Alice needs to investigate a missing taint flow (i.e., a bug arises at a sink, yet CodeQL does not issue a corresponding warning). She suspects that a third-party library is mistakenly modeled as a \emph{sanitizer}.
In Figure~\ref{fig:query-whynotflow}, the sensitive data travels from \texttt{user.getSSN()} through an external API: \texttt{format(SSN)} to \texttt{log(SSN)}.
After investigating the intermediate flow steps, Alice finds that \texttt{format(SSN)} was erroneously modeled as a sanitizer. 
% If this unreported path is unnoticed, a real security threat could have been undetected. 

\textsc{WhyNotFlow} visually identifies which APIs are acting as sanitizers (dashed arrows), pinpointing which API model could be responsible for killing a flow.
\textsc{WhyNotFlow} performs a speculative analysis by reasoning which taint flow path could have been {\em plausible} under a configuration where a sanitizer is instead modeled as a non-sanitizer.
In Figure~\ref{fig:query-whynotflow}, the graphical view marks the arrow after the node for  \texttt{format(SSN)} with a dashed line, indicating the flow currently does not exist, but would exist with a different model assumption. 

% if which third-party models were to change from a sanitizer to a non-sanitizer. 

\begin{figure}
    \centering
    \includegraphics[width=\linewidth]{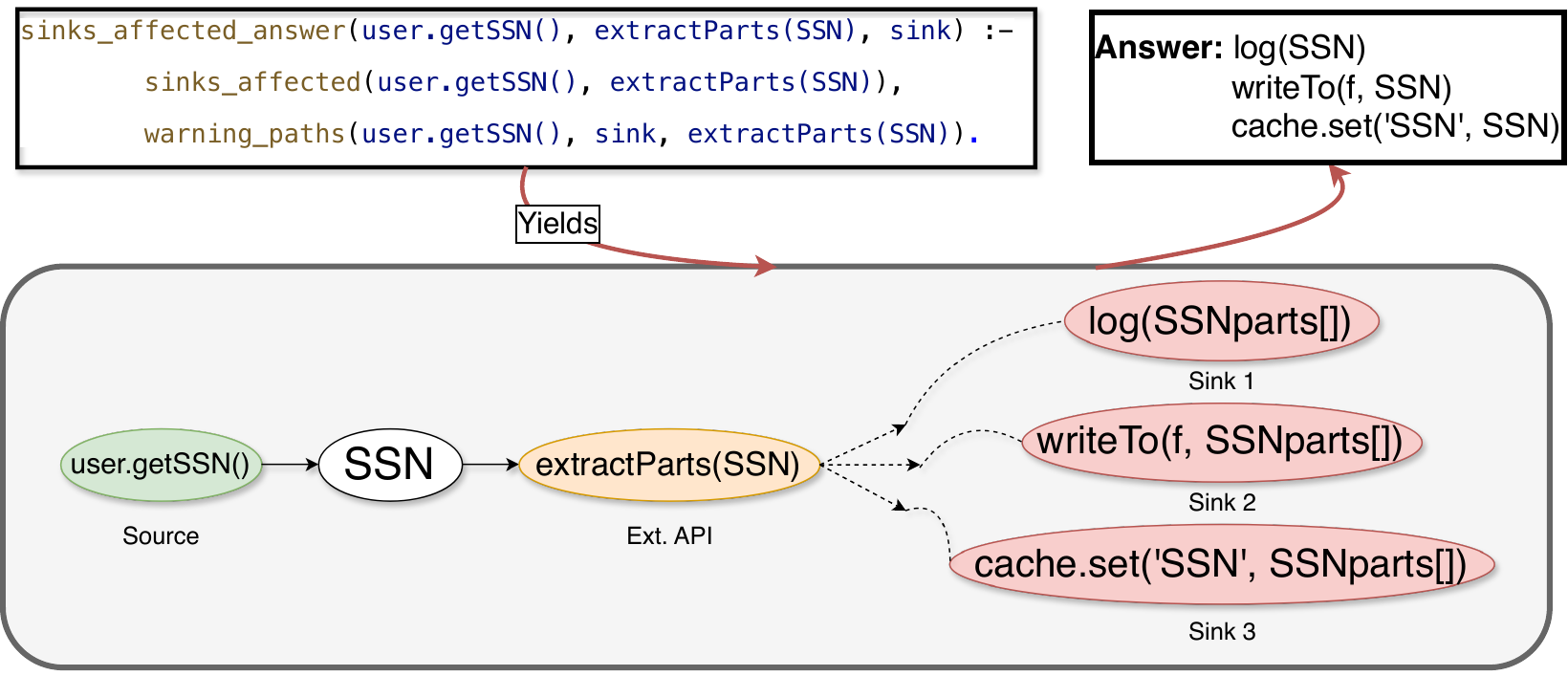}
    \caption{\textsc{AffectedSinks}: ``If we alter a third-party library's model, which sinks are affected?}
    \label{fig:query-affectedsinks}
\end{figure}

\noindent{\textbf{3. \textsc{AffectedSinks}: If we alter a third-party library $Z$'s model, which sinks are affected?}} Alice wants to reason about the global impact of updating a third-party library model~\cite{smith2015questions}. Suppose that she sees an unexpected warning and considers marking a third-party library's API as a sanitizer. However, she is concerned this update might suppress other warnings.
\emph{Without} \tool{}, she would need to sift through all reported warnings, manually trace each flow from the source, and check which sinks might become unreachable after her change\textemdash a time-consuming process.
\textsc{AffectedSinks} automates this what-if analysis, immediately revealing all \textcolor{red}{red} sinks that would be ``killed'' (Figure~\ref{fig:query-affectedsinks}).% Alice sees the effect of making \texttt{extractParts(SSN)} a sanitizer: all flows to sinks such as \texttt{log(SSN)}, \texttt{writeTo(f, SSN)}, and \texttt{cache.set(`SSN')} are terminated. By consolidating these results in one view, \tool{} spares her the effort of inspecting every warning individually, enabling a more strategic decision about whether the fix is too ``risky.''

\noindent{\textbf{4. \textsc{DivergentSinks} \& 5. \textsc{DivergentSources}:
Which third-party library model could influence multiple taint flows reaching the same sink (or originating from the same source)?}}

\begin{figure}
    \centering
    \includegraphics[width=\linewidth]{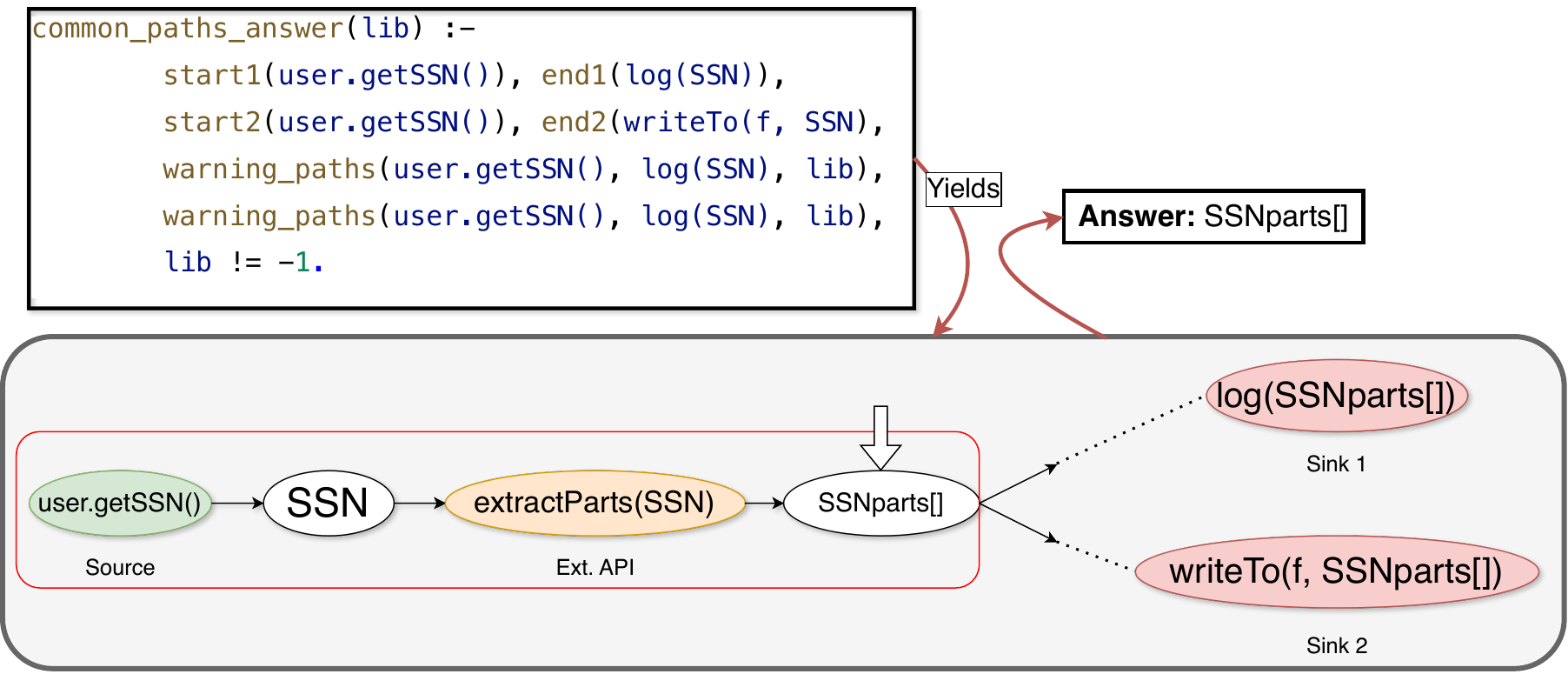}
    \caption{DivergentSinks: ``Which third-party library model could influence multiple taint flows from the same source?''}
    \label{fig:query-div-sinks}
%    \miryung{similarly, add a corresponding logic query and the answer}
\end{figure}

Suppose that Alice would like to know whether a known vulnerability reaching multiple \emph{sinks} could be fixed at once~\cite{smith2015questions,piskachev2019codebase}.
In a typical taint analyzer, it can be cumbersome to trace how a single piece of sensitive data, such as an \texttt{SSN}, propagates through multiple taint flows. Consequently, she may prefer to identify a common interception point instead of fixing one sink at a time. 
With \textsc{DivergentSinks} query, she can quickly locate a common point from the source that ``splits'' into multiple sinks. 
% If the branching point is from a third-party library function, she may choose to \emph{correct} the culprit model by reasoning about multiple taint flows at once 
(see Figure~\ref{fig:query-div-sinks}).  

To illustrate another scenario, imagine a single sensitive destination (sink) that receives data from multiple untrusted sources.
Using the \textsc{DivergentSources} query, Alice can readily identify if these diverse sources converge towards the same vulnerable point. This allows her to pinpoint the specific model responsible for the convergence, i.e., the `culprit model.'
Unlike traditional list-based warning views,
% which would force Alice to tediously compare individual data flow paths step-by-step,
\tool{} visually highlights where taint paths intersect, making it much easier to pinpoint the culprit model.

\begin{figure}
    \centering
    \includegraphics[width=\linewidth]{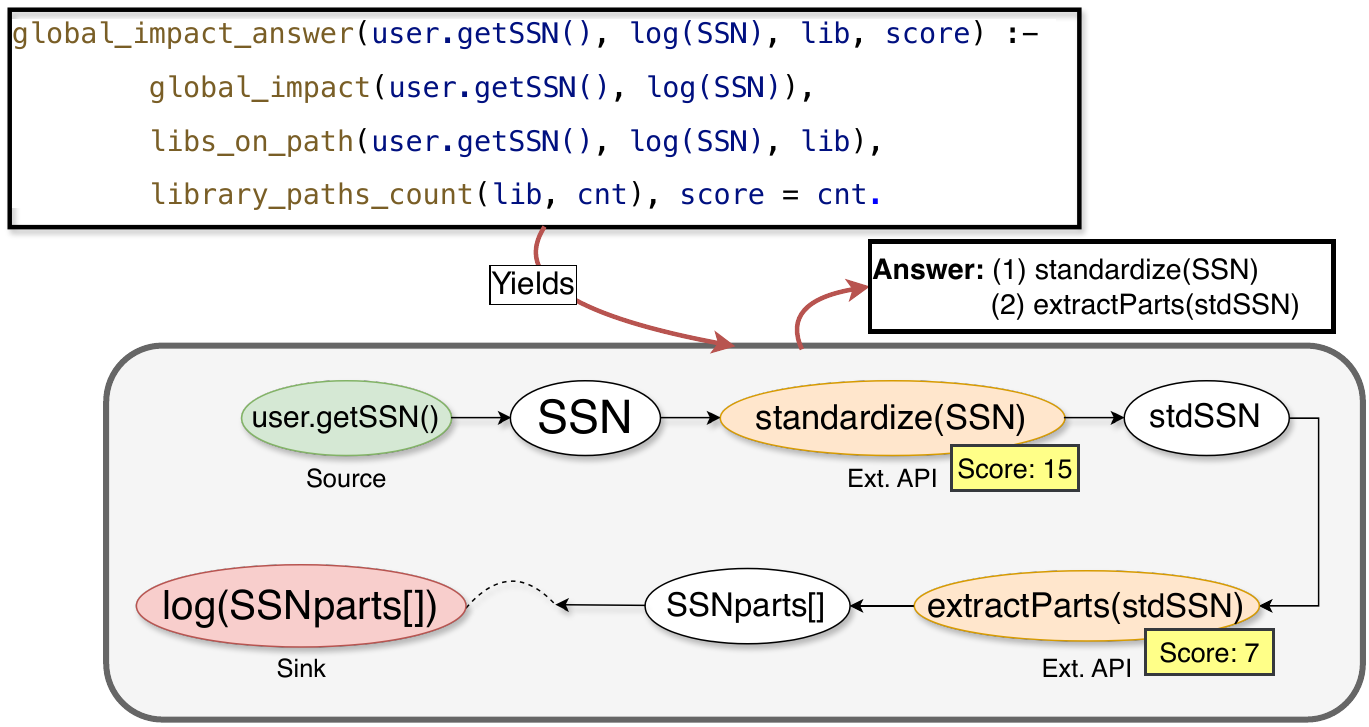}
    \caption{GlobalImpact: ``Which third-party library model could have the largest global influence on dataflows from a source to a sink?''}
    \label{fig:query-globalimpact}
\end{figure}

\noindent{\textbf{6. \textsc{GlobalImpact}: Which third-party library's model could have the largest global influence on multiple flows from a source to a sink?}}
When multiple models could explain a false or missing warning, developers like Alice often prefer a conservative fix that impacts the fewest flows~\cite{murphy2013design}. \emph{Without} \tool{}, identifying the frequency of each API in \emph{all} taint paths would require Alice to painstakingly review every single warning and count occurrences manually.
\textsc{GlobalImpact} automatically computes how often each API appears across multiple paths, ranking them by frequency. In Figure~\ref{fig:query-globalimpact}, because \texttt{extractParts(standardizedSSN)} appears in more paths, it has a higher score, signaling a larger global impact, which provides Alice with a better way to view the global impact of the APIs existing in the flows she is examining. 

\section{Speculative Analysis for Inquiry-based Debugging}\label{sec:approach}
\begin{figure*}
    \centering
    \includegraphics[height=0.44\textheight, width=\linewidth, keepaspectratio]{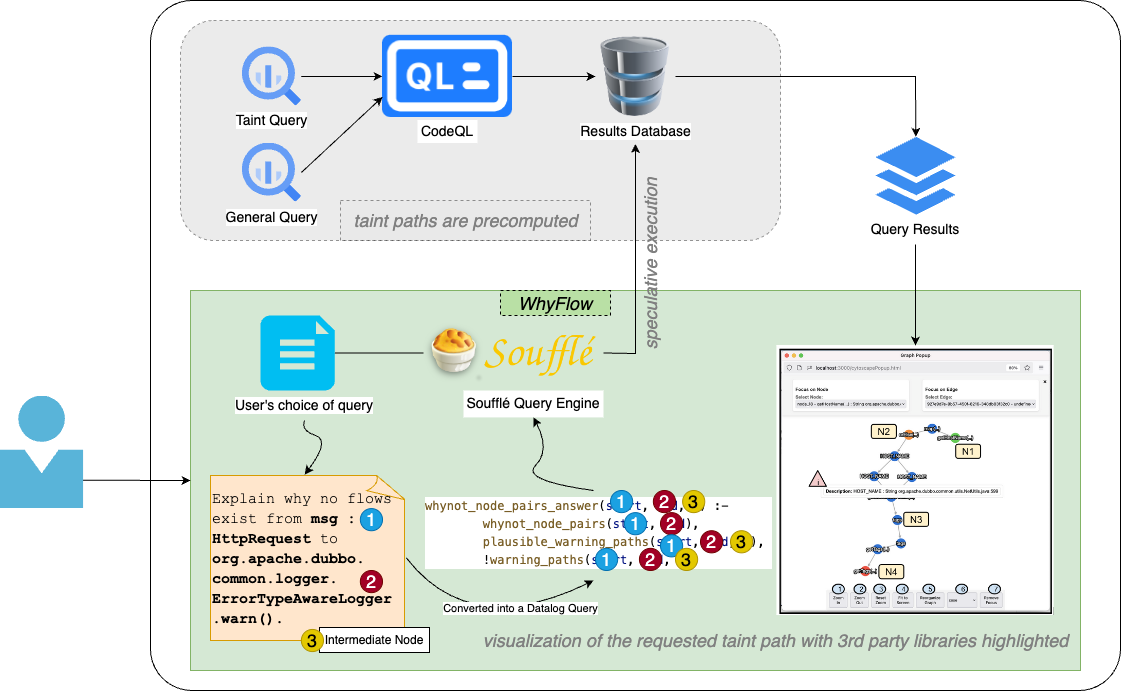}
    \caption{A user selects a \textit{WhyNotFlow} template query to investigate a suspected missing flow \texttt{msg\-:\-Http\-Request} to \texttt{Error\-Logger\-.warn()}. \tool{} concretizes a corresponding logic query. For instance, to explain a potential missing flow from \texttt{msg\-: Http\-Request} (node 1) to \texttt{Error\-Logger\-.warn()} (node 2), a plausible path via an intermediate node (node 3) is first identified by the rule \texttt{whynot\_node\_pairs\_answer(1,2,3)}. The query result is shown in the \textit{Graph View}.}
    \label{fig:whyflow-worklow}    
\end{figure*} 

%\hj{instead of "approach" as the section title, do you think we will be better off saying "Speculative Analysis for Inquiry-based debugging"?}

\tool{} enables inquiry-based debugging 
% for sensemaking to 
% debug 
of
spurious flows or missing flows.
Figure~\ref{fig:whyflow-worklow} shows the precomputation of the taint facts by executing both the original CodeQL taint query and a less restricted version of this query (Section~\ref{sec:codeql}).
After this, we convert the taint analysis results into \textit{Souffl\'e} facts and store them in a ``Results Database'' (Section~\ref{sec:souffle}).
Subsequently, \tool{} queries this database to provide answers to the template questions (Section ~\ref{sec:whyflow}).
% in \tool{}, 
% according to the user's choices (Section ~\ref{sec:whyflow}).
% \sout{.
% The answers are presented as an interactive graph visualization. In this section, we present the components of \tool{}, and how each of these components helps the overall workflow}.
% Before elaborating on our approach, 
We rely on several assumptions.
% in our design.

Assumption 1. We presume that a user of \tool{} is familiar with the program they are inspecting, meaning that they can recognize that some flows are unexpected flows and some expected flows are missing, serving as a starting point for end-user debugging.
% \sout{. This way, running an interrogative debugger can help narrow down the third-party library's model after selecting a relevant template query for `why' or `why-not' questions}.

Assumption 2. We leverage CodeQL to obtain taint analysis results as is, and then we utilize Souffl\'e to run the interrogative debugging queries on top of CodeQL results.
% \sout{. Hence, it is important to emphasize that CodeQL and Souffl\'e are being used together, and there is no modification to the underlying CodeQL's analysis algorithm. These two tools are not meant to compete}.

Assumption 3. We assume that CodeQL results are created with the 
information flow 
models of third-party libraries, which is typical for modern taint analysis.
We assume that inaccuracies in these plug-in models should be debugged, when a user suspects a missing flow or a spurious flow.
% \sout{. Since these models are plug-in abstractions and third-party libraries may not have source code, we assume that inaccuracies in these plug-in models should be debugged, when a user suspects a missing flow or a spurious flow}.

% \sout{Furthermore, t}
The goal of \tool{} is not to 
% frame `why' and `why-not' question as a 
disambiguate false positives from false negatives. 
% In fact, \tool{} never removes or adds any taint warnings reported by CodeQL. 
Rather, \tool{} helps with end-user debugging, explaining how configuration choices, such as source/sink definitions and third-party models, may lead to spurious or missing paths.
% Given that the target users need to maintain and update third-party models, \tool{} aims to simplifies this process.

\subsection{CodeQL}\label{sec:codeql}
CodeQL is an open-source static analysis framework. 
Users can use QL, which is a declarative, object-oriented Domain-Specific Language (DSL) specifically designed for writing CodeQL queries to detect potential security vulnerabilities or other issues~\cite{codeql_docs}. 
% \sout{.
% While CodeQL offers code scanning capabilities out of the box, its extendability via user-defined queries and models is particularly noteworthy}.

A key factor in CodeQL's comprehensiveness lies in its \emph{third-party library modeling}. Each library API (including third-party dependencies) can be modeled to reflect how data flows through its methods. 
% Recognizing that
As manual creation of such models is time-consuming and error-prone, the CodeQL team has explored 
% automated and 
machine-learning-based techniques to infer library behaviors \cite{nachtigall2022large}.
Nonetheless, incomplete or incorrect modeling remains a practical challenge.
% , especially in large ecosystems where libraries evolve frequently without thorough documentation of their input-output contracts.

%Beyond typical query use, developers sometimes run \emph{general} or \emph{speculative} queries in CodeQL—queries designed to enumerate all dataflows or suppress certain library-based assumptions. 
Rather than relying on specific third-party library models and source and sink definitions, a general query may relax these configuration assumptions, revealing the ``maximal'' set of reachable flows from all possible sinks to all possible sources, ignoring third-party library models. Such an approach can help users uncover paths that would otherwise be missed due to incomplete or inaccurate library modeling. Moreover, comparing results from both model-aware and ``maximal'' queries can expose discrepancies, hints that the current model assumptions may need refinement. Our work builds on these insights by introducing an interactive, interrogative debugger, where users can easily toggle assumptions about 
% sanitizers or 
third-party methods with pre-defined template queries.
%, moving from specific to general analysis without rewriting the underlying.

%In this study, we have repurposed one of the available queries that are hosted on the CodeQL repository on GitHub (Listing \ref{lst:untrusted-data}). \hj{This detail should go into Section 4 instead. In this section, we shouldn't present details specific to the user study. } This query looks for any data that can be passed to an external API, and reports these paths as warnings~\ref{}. The query enumerates the source and sink for each path, and the intermediary steps between them.
%\hj{This section provides background information about codeql, but instead of talking about how rules can be written by users, it may be more important to provide more details about how the models can be configured.}
%\hj{We should add background information about how CodeQL developers had a machine learning method for inferring models. Otherwise, a reader may really wonder about whether errors in models are a real problem.}
%\hj{Then, we should talk about how we perform the 'speculative' analysis by running the 'general' query. e.g.,  what is a 'general' query, why is it useful to run the 'general' query}

% \by{Connection with Fig of CodeQL UI}

\begin{figure}
    \centering
    \includegraphics[width=0.95\columnwidth]{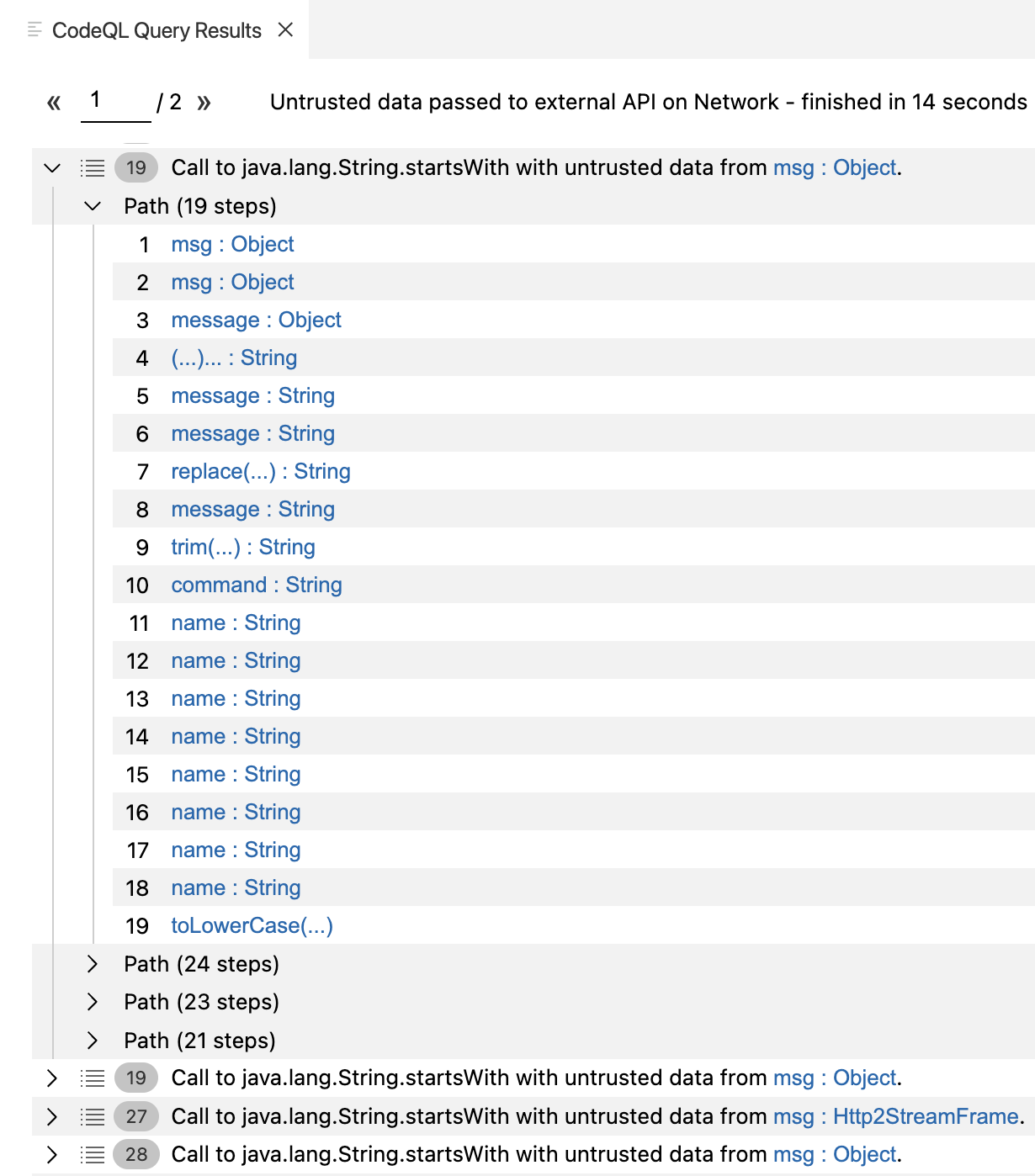}
    \caption{CodeQL's tree-based view shows each warning one by one and does not support inquiry-based sensemaking of multiple taint flows and the impact of user choices and third-party library models.
    % \hj{great addition! }
    }
    \label{fig:codeql-ui}
\end{figure}

\subsection{Souffl\'e and Logic Query Implementation}\label{sec:souffle}

\tool{} converts the CodeQL taint-analysis results into a set of \emph{facts} that can be analyzed by Souffl\'e's Datalog engine. In this factbase, each node, edge, and taint-related annotation (e.g., source, sink, and library API) is expressed as a logical predicate. Once the output of a single CodeQL query is transformed into these facts, \tool{} can quickly re-evaluate multiple ``secondary'' queries, such as those in Table~\ref{tab:queries}, without needing to re-run CodeQL's underlying taint analysis itself. 

This arrangement is particularly useful for tasks like missing-flow analysis (\textsc{WhyNotFlow}) or identifying global impact. Instead of incurring repeated analysis times of 10 to 14 seconds (excluding the compile time of the query) in CodeQL for each reconfiguration of the library model, we rely on Souffl\'e’s Datalog engine~\cite{jordan2016souffle} to query the precomputed facts in the order of 5 seconds---a speed-up that supports interactive debugging. 
% Hence, we selected Souffle~\cite{} for its ease of rapid prototyping, and its high performance and scalability in developing static analysis~\cite{}. 

\paragraph{Converting the CodeQL Output to Facts.}
Each CodeQL dataflow node is mapped to a \texttt{node(\emph{id})} predicate, capturing its unique identifier and associated metadata (e.g., filename, line/column, symbol name).
Likewise, each dataflow edge is encoded as \texttt{edge(\emph{edgeid}, \emph{sourceid}, \emph{targetid})}, indicating potential taint propagation.
We also store ``plausible edges''---those that might be inactive under certain sanitizers or library assumptions---using a similar \texttt{plausible\_edge} predicate. 
Sources and sinks become \texttt{source(\emph{nodeid})} and \texttt{sink(\emph{nodeid})}, while known library-flow relations (e.g., which arguments flow into return values) are recorded with \texttt{library\_flow(\emph{edgeid}, \emph{fact\_id})}. 

\paragraph{Answering \textsc{``WhyFlow''} and \textsc{``WhyNotFlow''}}
Once loaded into Souffl\'e, the questions in Table~\ref{tab:queries} are encoded as logic queries shown in Figures \ref{fig:query-whyflow} and \ref{fig:query-whynotflow}. For example, \textsc{WhyFlow} query uses transitive closure over \texttt{edge} to find whether data can reach a sink from a source, intersecting with \texttt{library\_flow} to highlight which third-party library's APIs (orange nodes) appear along the path.
%\begin{lstlisting}[basicstyle=\small\ttfamily]
%reachable(Src, Snk) :- 
%  source(Src),
%  sink(Snk),
%  edge(_, Src, Mid),
%  ... // transitive closure
%  edge(_, Prev, Snk).
%intermediateAPI(A) :-
%  reachable(Src, Snk),
%  library_flow(EdgeID, FactID),
%  // unify 'EdgeID' with edges that appear 
%  // in the path from Src to Snk
%  ...
%\end{lstlisting}
\textsc{WhyNotFlow} similarly checks for ``broken'' transitive paths and identifies \texttt{library\_flow} edges that act as sanitizers, which terminate flows.

\paragraph{Supporting \textsc{``DivergentSources/Sinks''} and \textsc{``GlobalImpact''}.}
\textsc{DivergentSinks} and \textsc{DivergentSources} queries use logic rules to detect the last common node or first common node in two paths. 
They compute the intersection of reachable sets for each source/sink pair. 
Meanwhile, \textsc{GlobalImpact} queries count how frequently an API appears in distinct \texttt{reachable} flows; each API node is assigned a score via an aggregation rule.
% over the factbase. 
\tool{} sizes the node (in the visualization) based on its score, visualizing its “global impact.”

\subsection{\tool{}'s User Interface}\label{sec:whyflow}
To support sensemaking, key design elements in \tool{} include:

\paragraph{Color‐Coding and Visual Clarity.}
Nodes in the flow graph are color‐coded:
\textcolor{green}{\textbf{green}} for sources,
\textcolor{red}{\textbf{red}} for sinks,
\textcolor{orange}{\textbf{orange}} for external APIs or libraries,
and \textcolor{blue}{\textbf{blue}} for other intermediate nodes.

\paragraph{Graphical Flows and Expandable Paths.}
\tool{} overlays the taint flows onto a single graph.
Solid edges represent dataflow steps.
Dashed edges indicate ``plausible flows'' currently unreported by the analyzer, but can be reported under a different configuration.
Each flow can be expanded to narrow down onto suspicious segments.

\paragraph{Clickable Nodes and Code Navigation.}
Clicking on a node opens the corresponding code snippet in the user's IDE for easy reference of the source code.
% making it easier to confirm ambiguous method names or library calls.
\tool{} includes hover popups showing details such as fully-qualified names of identifiers referenced by the intermediate steps along the flow.

\paragraph{Customizable Layouts and Node Sizing.}
\tool{} supports multiple layout algorithms (e.g., breadth‐first, concentric).
% \sout{.
% Users can reorganize the graph
% library‐heavy sections
% for improved clarity}.
A \emph{GlobalImpact} query resizes API nodes by their number of occurrences on different flows.

\section{User Study}\label{sec:userstudy}
To evaluate {\tool}'s usefulness in sensemaking taint flows, we designed a within-subject study. We assess how users can reason about the impact of user configurations on multiple taint flows, including third-party taint analysis models. We use CodeQL's Visual Studio Code plugin (CodeQL Visualizer in short), as the baseline. 
\paragraph{Study Design}
Participants were asked to answer eight sensemaking questions about taint analysis results and how the configuration of third-party libraries impacts taint flows. Each user study task consists of eight questions shown in Table~\ref{tab:programpoints}. The first six questions are centered on why, why-not, and what-if questions about taint flows and the remaining two are focused on quantifying taint flows. 

\paragraph{Research Questions}
\begin{enumerate}
    \item How much does \tool{} improve the participants' ability to answer questions about the configuration's impact on taint flows?
    \item How does \tool{} influence cognitive load and user confidence in sensemaking taint flows?
    \item What are the participants' perceptions of \tool{}'s usability and functionality in enhancing their workflow?
\end{enumerate}

\begin{table}[htbp]
    \centering
    \begin{tabular}{lc}
        \toprule
        \textbf{Metric} & \textbf{Value} \\
        \midrule
        \# of edges & 6,901 \\
        \# of nodes & 8,101 \\
        \# of sources & 26 \\
        \# of sinks & 265 \\
        \# of third-party API functions & 85 \\
        %Size of logic facts & xxx \\
       % Time taken to run query & xxx \\
        \bottomrule
    \end{tabular}
    \caption{Statistics of Taint Analysis Facts for \texttt{Apache Dubbo~\cite{Kevin2025apache}}. %\by{There is a comment by the reviewer saying that "Table 2 is not directly visible". I don't quite get what they mean by that.}
    }
    \label{tab:facts}
\end{table}

\subsection{Study Protocol}\label{sec:study-protocol}

The study task is based on 383 taint warnings generated on Apache Dubbo~\cite{Kevin2025apache} with the CodeQL query shown in Listing \ref{lst:untrusted-data}~\cite{codeql_externalapi_query}. The resulting facts from this query are presented in Table~\ref{tab:facts}.
% This study involves examining 383 taint flow warnings. %We chose a subset of 16 paths for two question sets containing the questions in Table~\ref{tab:queries} for our user study. 

\begin{listing}
\caption{CodeQL Taint Analysis Query~\cite{codeql_externalapi_query}.}
\label{lst:untrusted-data}
\noindent
\begin{lstlisting}[
  language=Java,
  numbers=left,           % Line numbers
  breaklines=true,        % Wrap long lines
  breakatwhitespace=false,% Allow breaking anywhere
  basicstyle=\ttfamily\small, % Smaller font size
  xleftmargin=0pt,        % No extra left margin
  frame=lines             % Frame around code
]
import java
import semmle.code.java.dataflow.FlowSources
import semmle.code.java.dataflow.TaintTracking
import semmle.code.java.security.ExternalAPIs
import UntrustedDataToExternalApiFlow::PathGraph

from UntrustedDataToExternalApiFlow::PathNode source, 
     UntrustedDataToExternalApiFlow::PathNode sink
where UntrustedDataToExternalApiFlow::flowPath(source, sink)
select sink, source, sink,
  "Call to " + sink.getNode().(ExternalApiDataNode).getMethodDescription() +
  " with untrusted data from $@.", source, source.toString()
\end{lstlisting}
\end{listing}

\subsubsection{Baseline}
We selected the CodeQL plugin in VSCode (CodeQL visualizer in short), shown in Figure \ref{fig:codeql-ui} as our baseline tool. It provides an interface where warnings are listed one-by-one, grouped under each analysis kind. Users can click and expand each warning to reveal the flows that contributed to the warning. Each flow can be expanded to show the steps in the flow. A user inspects warnings one-by-one.

\subsubsection{Participants}
We conducted a within-subject user study with a crossover design. We recruited 12 participants, including graduate students and professional developers from the industry. Their programming experience ranged from 1--3 years (4 participants), 4--6 years (3 participants), 7--10 years (3 participants), to over 10 years (2 participants). The average self-reported familiarity with taint analysis was 2.1 out of 5, suggesting that most participants had only moderate experience with this type of dataflow analysis. 
% As our study design includes a crossover, which minimizes variability, we require fewer subjects~\cite{vegas2015crossover} than a between-subject study.
%\by{We should delete this last sentence.}
%We recruited 13 participants by reaching out to students in Computer Science at various levels and with different years of experience. 

7 participants are PhD students, 3 are MS students, 1 is an undergraduate and 1 is a professional developer from industry. We dropped one participant from the study, since the participant failed to complete half of both tasks.
%\hj{are the numbers right? should they add up to 13?} \by{We dropped one of the participants}
%\hj{oh, we should write this in the paper and explained why we dropped them}
%\hj{TODO for Burak}
%The average self-reported familiarity with taint analysis was 2.1 out of 5. 
% As the investigation of a large collection of warnings is task that requires significant cognitive engagement,
% we hypothesized that the participant's information processing and need for cognition may impact their behavior and use of \tool{}.
% As such, we collected these additional pieces of information and analyzed how they influenced the participants' inspection strategies.

\paragraph{Number of participants.} While between-subject user studies require a large number of participants to account for variations among individual participants, within-subject user studies minimize variability, as each participant uses both tools following a different order. 
The order of tool usage and task assignment is randomized and counterbalanced~\cite{vegas2015crossover, chasins_2021}. 
Within-subject user studies with 8 to 16 participants are standard practice in both software engineering~\cite{garcia_2024, kang_2024, ganji_2023} and HCI research~\cite{horvath_2022, suh_2023, huh_2024}.

\subsubsection{Protocol}
Each participant took part in a 1.5-hour session. The study involved using both \tool{} and the baseline CodeQL visualizer for the two tasks. 
The order of assigned tool (\tool{} first vs.\ CodeQL first) and the assigned task (Problem Set~\#1 and Problem Set~\#2) was counterbalanced across participants through random assignment. We gave a 3-minute pre-study survey to collect background information, followed by a tutorial.

Next, each participant proceeded with the two tasks, each lasting 20 minutes. 
Each task had 8 questions that participants were tasked to answer. They had the option of skipping over questions and ending the task early. 
After each task, participants filled out a 5-minute post-task survey, which included the NASA-TLX questions~\cite{hart1988development} (e.g., mental demand, time pressure, perceived success, effort, and frustration). 
In the final 12-minute post-study survey, we asked participants to compare \tool{} and CodeQL, and rate \tool{}'s features.

\subsubsection{Tutorial}
We conducted a tutorial session to introduce participants to the functionality of both \tool{} and the CodeQL VSCode plugin.
% The tutorials were structured around realistic dataflow scenarios.
For \tool{}, we guided users through all six query types.
% (\emph{Why\-Flow}, \emph{Why\-Not\-Flow}, \emph{Affected\-Sinks}, \emph{Divergent\-Sinks}, \emph{Divergent\-Sources}, and \emph{Global\-Impact}).
This involved selecting relevant sources, sinks, or API calls from dropdown menus and interpreting the graphical output rendered by \tool{}.
% Participants were encouraged to hover over nodes to view fully qualified names, click on nodes to navigate directly to code, and adjust the layout using built-in options.
Participants were encouraged to interact with \tool{}'s features to familiarize themselves with them. 
% Through these interactions, participants familiarized themselves with \tool{}'s features. 

% , ensuring that even those unfamiliar with taint analysis could follow each step. 
% The tutorial format allowed us to illustrate the advantages of visual interrogation, such as quickly locating divergent flows, comparing paths, and gauging the “global impact” of particular APIs. This grounding also ensured that participants understood the motivation behind each query type, promoting more consistent use of \tool{} during later study tasks. 

We also conducted a tutorial session on using CodeQL's visualizer.
% Participants learned how the plugin displays query results in a tabular list.
% We walked them through the process of expanding or collapsing the flows presented in the visualizer to inspect the intermediate steps between source and sink.
We walked the participants through the process of viewing query results in a tabular list, and expanding or collapsing the flows to inspect the intermediate steps between source and sink.

Finally, we explained the role of CodeQL ``models'' of external libraries, highlighting their role in the analysis.
% \sout{.
We reinforced core concepts—sources, sinks, and sanitizers—and provided examples.
% This hands-on demonstration ensured that participants understood both CodeQL’s strengths—such as its detailed path expansions and direct code navigation—and its limitations with respect to external library assumptions, preparing them to perform study tasks and compare CodeQL’s interface with \tool{}. 

\subsubsection{Tasks}
Each task had a different set of questions (Problem Set~\#1 and Problem Set~\#2). 
The questions and their answers are shown in Table~\ref{tab:programpoints}. 
%To minimize learning effects, we balanced the order of tools that participants used first. 
Each task contained eight questions about `why', `why-not', `what-if,' as shown in Table~\ref{tab:queries}, and two additional questions about the quantitative aspects of taint flows (e.g., how many taint paths exist or how many third-party APIs appear in the taint paths). The questions were multiple-choice questions, some having multiple answers. We granted partial credit if participants selected some correct answers, but missed or added incorrect choices.

% We designed these tasks to reflect realistic inquiries developers make when investigating taint-analysis results in large projects. Specifically, the questions ask participants to identify which third-party library calls sanitize or propagate data, locate shared pass-through points for merging or diverging paths, and measure the overall magnitude of taint propagation. In creating these two problem sets, we ensured coverage of diverse scenarios: flows with multiple third-party APIs, flows requiring path comparisons, and flows that might easily be missed by a baseline listing of warnings.\by{I can remove this paragraph if you think it is redundant.}

% We presented the tasks in a form-based interface, providing multiple-choice or short-answer prompts. For instance, \textsc{AffectedSinks} questions required participants to select all sinks that would be eliminated if a particular API were modeled as a sanitizer, while \textsc{TraceLens} required naming the intermediate library call that enabled a flow from a known source to a known sink. The final two “quantitative” questions in each set asked for the number of pass-through points in a single flow or the total number of distinct flows leading from a given source to a sink.

% \hj{should we provide more details about the problem sets?}
% \hj{We should provide more information about the problem sets. What sort of questions do we ask? Why did we pick these questions?}\by{Done?}

% \hj{I skipped the details of the problem set}\by{table done}
%Below, we list the contents of each problem set in detail.

{\footnotesize
\begin{table*}[!htbp]
\centering
\resizebox{\textwidth}{!}{%
\fbox{\begin{minipage}{0.98\textwidth}
\rowcolors{2}{gray!10}{white}
\renewcommand{\arraystretch}{1.3}
\begin{tabular}{p{0.48\textwidth}p{0.48\textwidth}}
\toprule
\textbf{Problem Set \#1} & \textbf{Problem Set \#2} \\
\midrule
\begin{minipage}[t]{\linewidth}
(1) Explain why a taint flow is permitted from \texttt{getRequestURI(...)} in \texttt{[...].PageServlet.java} to charAt(...) in \texttt{[...].StringUtils.java}. Name a third-party API permitting the flow.\\
\textbf{Answer: } \texttt{java.lang.String.substring(int beginIndex, int endIndex)}
\end{minipage} &
\begin{minipage}[t]{\linewidth}
(1) Explain why a taint flow is permitted from msg: String in \texttt{[...].TelnetProcessHandler.java} to \texttt{json: String} in \texttt{[...].FastJsonImpl.java}. Name a third-party API permitting the flow.\\
\textbf{Answer: } \texttt{java.lang.String.substring(...)}
\end{minipage} \\
\midrule
\begin{minipage}[t]{\linewidth}
(2) Explain why no taint flow is permitted from \texttt{msg: HttpRequest} to \texttt{warn(...)} in \texttt{[...].HttpProcessHandler.java}.\\
\\
\textbf{Answer: } \texttt{io.netty.handler.codec.http.HttpRequest.method(...)}
\end{minipage} &
\begin{minipage}[t]{\linewidth}
(2) Explain why no taint flow is permitted from \texttt{msg: Http2StreamFrame} in \texttt{[...].TripleHttp2ClientResponseHandler.java} to release(...) in \texttt{[...].TripleClientStream.java}.\\[1ex]
\textbf{Answer: } \texttt{io.netty.handler.codec.http2.}
\texttt{Http2DataFrame.isEndStream()}
\end{minipage} \\
\midrule
\begin{minipage}[t]{\linewidth}
(3) Explain what sinks would no longer be reachable, if \texttt{io.netty.handler.codec.http.HttpRequest.uri()} were modeled as a sanitizer, starting from source \texttt{msg: HttpRequest} in \texttt{[...].HttpProcessHandler.java}.[1ex]
\\
\textbf{Answer: } \texttt{valueList} in \texttt{[...].HttpCommandDecoder.java} and \texttt{msg} in \texttt{[...].Log4jLogger.java}
\end{minipage} &
\begin{minipage}[t]{\linewidth}
(3) Explain what sinks would be no longer reachable, if \texttt{io.netty.handler.codec.http2.Http2HeadersFrame.headers()} is marked as a sanitizer, starting from source \texttt{msg: Object} in \texttt{[...].TripleHttp2FrameServerHandler.java}.\\[1ex]
\textbf{Answer: } path: \texttt{[...].TriplePathResolver.java} and headers: \texttt{[...].TripleIsolationExecutorSupport.java}
\end{minipage} \\
\midrule
\begin{minipage}[t]{\linewidth}
(4) Identify the program point that affects multiple taint flows ending at two sinks: 
\\[0.5ex]
\quad \textit{sinks:} \texttt{path: String} in \texttt{[...].TriplePathResolver.java} at line 41 and \texttt{path: String} in \texttt{[...].TriplePathResolver.java} at line 46, \textit{source:} \texttt{msg: Http2StreamFrame} in \texttt{[...].TripleHttp2ClientResponseHandler.java}.\\
\textbf{Answer: } \texttt{toString(...): String in [...].TripleServerStream.java}
\end{minipage} &
\begin{minipage}[t]{\linewidth}
(4) Identify the program point that affects multiple taint flows ending at two sinks: 
\\[0.5ex]
\quad \textit{sinks:} \texttt{path: String} in \texttt{[...].TriplePathResolver.java} – line 41 and \texttt{path: String} in \texttt{[...].TriplePathResolver.java} – line 46, \textit{source:} \texttt{msg: Object} in \texttt{[...].TripleHttp2FrameServerHandler.java}.\\
\textbf{Answer: } \texttt{toString(...): String in [...].TripleServerStream.java}
\end{minipage} \\
\midrule
\begin{minipage}[t]{\linewidth}
(5) Identify the intermediary program point that influences multiple flows originating from \texttt{input: ByteBuf} in \texttt{[...].NettyCodecAdapter.java} and in: ByteBuf in \texttt{[...].NettyPortUnificationServerHandler.java}, ending at \texttt{buffer: ByteBuf} in \texttt{[...].NettyBackedChannelBuffer.java}.\\
\\
\textbf{Answer: } \texttt{parameter this : NettyBackedChannelBuffer [buffer]: ByteBuf in [...].NettyBackedChannelBuffer.java}
\end{minipage} &
\begin{minipage}[t]{\linewidth}
(5) Identify the intermediary program point that influences multiple flows originating from \texttt{msg: Http2StreamFrame} in \texttt{[...].TripleHttp2ClientResponseHandler.java} and \texttt{msg: Object} in \texttt{[...].TripleHttp2FrameServerHandler.java}, ending at \texttt{path: String} in \texttt{[...].TriplePathResolver.java}.\\
\textbf{Answer: } \texttt{headers: Http2Headers in [...].TripleServerStream.java}
\end{minipage} \\
\midrule
\begin{minipage}[t]{\linewidth}
(6) Which third-party APIs could have the most influence on the taint path from \texttt{msg: Object} in \texttt{[...].TripleHttp2FrameServerHandler.java} to \texttt{path: String} in \texttt{[...].TriplePathResolver.java}? Rank in the order of importance.\\
\textbf{Answer: }\\ 
\texttt{(1) io.netty.handler.codec.http2.Http2HeadersFrame.headers()}\\
\texttt{(2) java.lang.CharSequence.toString()}\\ 
\texttt{(3) io.netty.handler.codec.http2.Http2Headers.path()}
\end{minipage} &
\begin{minipage}[t]{\linewidth}
(6) Which third-party APIs could have the most influence on the taint path from \texttt{msg: Object} in \texttt{[...].NettyClientHandler.java} to \texttt{key: String} in \texttt{[...].TraceFilter.java}? Rank in the order of importance.\\
\textbf{Answer: }\\
\texttt{(1) java.lang.String.trim()}\\
\texttt{(2) java.lang.String.replace(...)}\\
\texttt{(3) java.lang.String.substring(...)}
\end{minipage} \\
\midrule
\begin{minipage}[t]{\linewidth}
(7) Determine the number of pass-through API points from \texttt{getRequestURI(...)} in \texttt{[...].PageServlet.java} to \texttt{charAt(...)} in \texttt{[...].StringUtils.java}.\\
\\
\textbf{Answer: } \texttt{21}
\end{minipage} &
\begin{minipage}[t]{\linewidth}
(7) Determine the number of pass-through API points from \texttt{msg: Http2StreamFrame} in \texttt{[...].TripleHttp2ClientResponseHandler.java} to \texttt{path: String} in \texttt{[...].TriplePathResolver.java}.\\
\textbf{Answer: } \texttt{21}
\end{minipage} \\
\midrule
\begin{minipage}[t]{\linewidth}
(8) Count how many different dataflow paths exist from \texttt{in: ByteBuf} in \texttt{[...].NettyPortUnificationServerHandler.java} to \texttt{buffer: ByteBuf} in \texttt{[...].NettyBackedChannelBuffer.java}.\\
\textbf{Answer: } \texttt{2}
\end{minipage} &
\begin{minipage}[t]{\linewidth}
(8) Count how many different dataflow paths exist from \texttt{msg: Http2StreamFrame} in \texttt{[...].TripleHttp2ClientResponseHandler.java} to \texttt{path: String} in \texttt{[...].TriplePathResolver.java}.\\
\textbf{Answer: } \texttt{4}
\end{minipage} \\
\bottomrule
\end{tabular}
\end{minipage}}%
}%
\caption{Study Tasks: Problem Sets \#1 and \#2 with their correct answers. Participants were assigned one of the two tasks to complete with CodeQL visualizer and the other with \tool{}.}
\label{tab:programpoints}
\end{table*}
}

\subsubsection{Criteria for Study Tasks}

% In order to avoid a study task design where participants can directly gather answers using \tool{}'s QA interface, we purposely designed questions to require reasoning of multiple flows and multiple third-party APIs, and reasoning non-existing flow paths that would have existed under a different model assumption.
We designed questions based on realistic scenarios involving the analysis of taint flows.
% from sources to sinks
These questions correspond to ``why'', ``why-not'', and ``what-if'' questions that developers may ask about a taint analysis or assess potential changes to models of the external libraries.

\paragraph{Post-Study Questionnaire.}
After completing each task, participants completed a brief post-task questionnaire, which included the NASA-TLX questions. They also described their strategies to find answers to the questions in this questionnaire. At the end of the session, participants rated user-interface features, confidence, and ease of use.
% with both tools.  

%\begin{table}[h!]
%\centering
%\setlength{\tabcolsep}{4pt} % Adjust column spacing
%\renewcommand{\arraystretch}{1.2} % Adjust row height
%\begin{tabular}{|p{0.5cm}|p{1.8cm}|p{1.8cm}|p{1.5cm}|p{1.2cm}|}
%\hline
%\textbf{ID} & \textbf{Task 1} & \textbf{Task 2} & \textbf{Problem Set 1} & \textbf{Problem Set 2} \\ \hline
%1  & CodeQL  & TraceLens & 1 & 2 \\ \hline
%2  & TraceLens & CodeQL  & 1 & 2 \\ \hline
%5  & CodeQL  & TraceLens & 2 & 1 \\ \hline
%4  & TraceLens & CodeQL  & 2 & 1 \\ \hline
%3  & CodeQL  & TraceLens & 1 & 2 \\ \hline
%6  & TraceLens & CodeQL  & 1 & 2 \\ \hline
%7  & CodeQL  & TraceLens & 2 & 1 \\ \hline
%8  & TraceLens & CodeQL  & 2 & 1 \\ \hline
%9  & CodeQL  & TraceLens & 1 & 2 \\ \hline
%10 & TraceLens & CodeQL  & 1 & 2 \\ \hline
%11 & CodeQL  & TraceLens & 2 & 1 \\ \hline
%12 & TraceLens & CodeQL  & 2 & 1 \\ \hline
%\end{tabular}
%\caption{Task Assignments and Problem Sets for Participants \hj{we don't have to include this table in the paper} \by{BURAK TODO: Remove this table}}
%\label{tab:task_assignments}
%\end{table}

\section{Results}\label{sec:results}
\begin{figure}
    \centering
    \includegraphics[width=0.95\columnwidth]{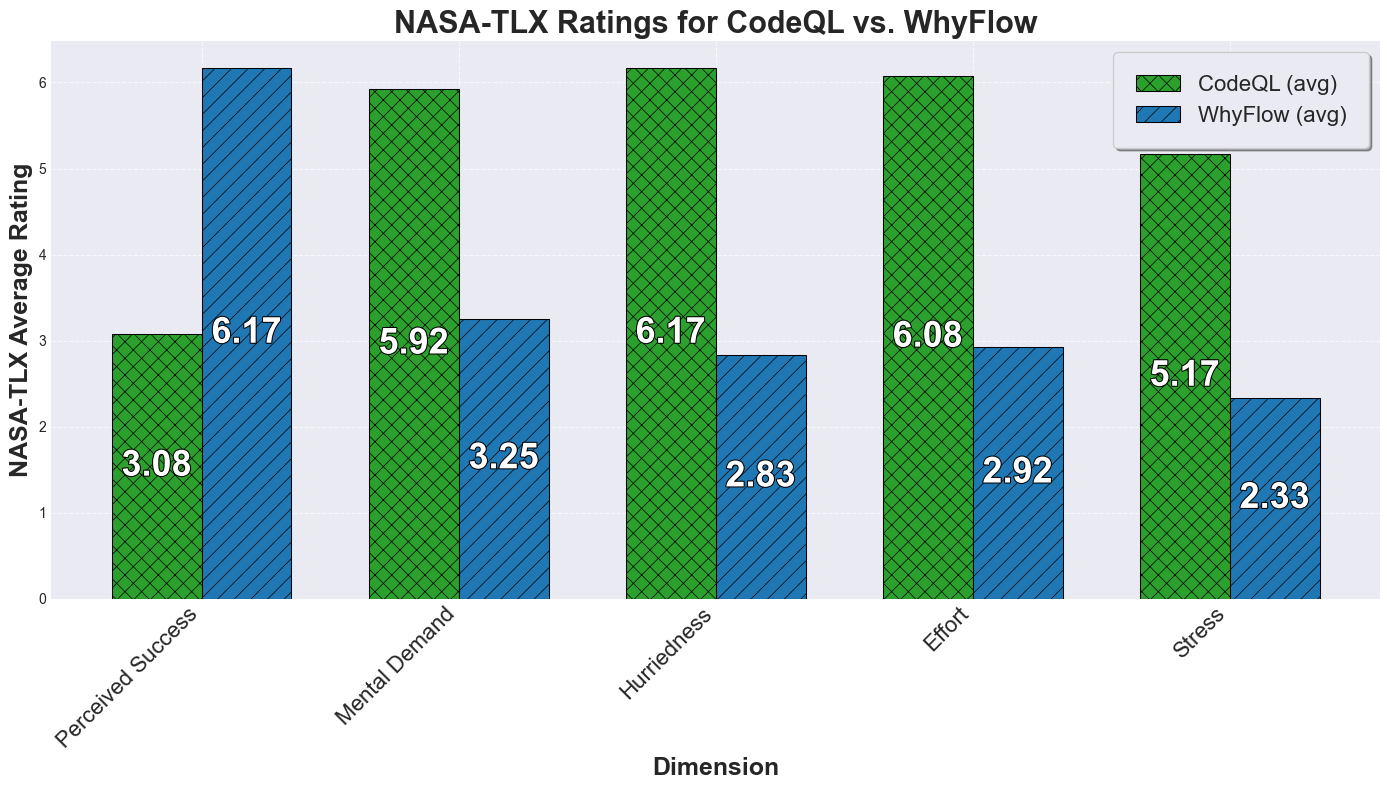}
    \caption{Average NASA‑TLX Ratings for \tool{} vs. CodeQL visualizer. %This figure reports the mean NASA‑TLX scores across five workload dimensions: Mental Demand, Hurriedness, Perceived Success, Effort, and Stress. Lower scores are better for each dimension other than Perceived Success, where higher values are better. 
   % Compared to CodeQL, \tool{} generally led to a lower perceived workload, including less cognitive demands and stress, along with higher perceived success. 
    % \hj{the poor color contrast makes it hard to read the graph. The text is also a little too small}\by{Done.}
    }
    \label{fig:nasa-tlx}
\end{figure}

%Both participants reported lower mental demand and stress when using TraceLens, alongside higher perceived success. They also indicated greater confidence in their answers with TraceLens (5 out of 5) than with CodeQL (2--3 out of 5).
In this section, we analyze the results of our user study. 
%Our study follows a within-subject crossover design where each participant used both \tool{} and CodeQL on different tasks.
In our evaluation, use non-parametric statistical tests because our data is inherently ordinal, including Likert-scale responses (NASA-TLX ratings from 1--7, confidence ratings from 1--5, ease-of-use ratings from 1--5) and categorical accuracy assessments (correct, partial, incorrect, empty).
Specifically, we apply the two-sided \emph{Mann-Whitney U test}~\cite{Ashby_1991, Student_1908} to compare distributions between \tool{} and CodeQL, as is standard practice for ordinal data in user studies~\cite{bavota_2013, yu_2024, fast_2018, liu_2024, warford_2021}.
We report statistical significance at the $\alpha = 0.05$ level.
We also compute Cohen's $d$ to measure effect size using the pooled standard deviation, where $|d| > 0.8$ indicates a large effect~\cite{cohen1988statistical}.

We test the following null hypotheses:
\begin{itemize}[leftmargin=*,noitemsep,topsep=0pt]
    \item \textbf{RQ1 (Accuracy):} $H_0$: There is no difference in the distribution of participants' accuracy (correct, partial, or empty responses) between \tool{} and CodeQL.
    \item \textbf{RQ2 (Cognitive Load):} $H_0$: There is no difference in the distributions of participants' 
    % self-reported 
    mental demand, hurriedness, perceived success, effort, or stress between \tool{} and CodeQL.
    \item \textbf{RQ3 (Usability):} $H_0$: There is no difference in the distributions of participants' confidence or ease-of-use ratings between \tool{} and CodeQL.
\end{itemize}
We reject each null hypothesis when $p < 0.05$, indicating that observed differences are statistically significant and unlikely due to chance.
Each participant is denoted as P\#. 

%We also report standard deviations for accuracy metrics (RQ1) and NASA-TLX ratings (RQ2) to capture variability in responses.
% \by{Text update about statistical tests}
% \by{Update the reference in the replication package}

\subsection{RQ1. Accuracy}

Table~\ref{tab:tracelens_accuracy} shows the participants' accuracy in answering questions for the problem sets (\emph{Task~1} and \emph{Task~2}) when using \tool{} and the CodeQL Visualizer.
% Table~\ref{tab:tracelens_accuracy} shows the accuracy of answers for across \emph{Task~1} and \emph{Task~2} when using \tool{} and CodeQL visualizer, respectively.
Accuracy outcomes showed a clear advantage for \tool{}, where users had more correct answers than using the CodeQL visualizer ($p < 0.05$), alongside fewer empty responses ($p < 0.05$). % U-test $p < 0.0099$, t-test $p = 0.0305$

These differences are statistically significant according to the Mann-Whitney U test with a large effect size (\emph{Correctness:} Cohen's $d = 1.33$, \emph{Empty answers:} Cohen's $d = 1.8$).
We therefore reject the null hypothesis for RQ1: \tool{} significantly improves accuracy compared to CodeQL.
Participants using \tool{} provided more correct answers, especially when needing to reason about multiple flows and global impact; on the other hand, participants using the CodeQL visualizer felt overwhelmed~\cite{schwartz2015paradox}.
% \hj{TODO; fill in}).
%In \emph{Task 2}, the standard deviation for correct answers was lower in TraceLens ($std. dev. = 1.23$) than in CodeQL ($std. dev. = 1.70$), indicating both higher and more consistent performance. 
%\hj{If we added the paragraph about the choice of statistical tests at the beginning of Section 5, then perhaps we don't have to repeat "Mann-Whitney U Test"}

%In \emph{Task 1}, while TraceLens users again achieved higher correct answer counts, the difference did not reach statistical significance (t-test $p = 0.1248$; U-test $p = 0.1030$). 
%However, TraceLens still led to significantly fewer empty responses in \emph{Task 1} (t-test $p = 0.0027$; U-test $p = 0.0022$).
% , with zero standard deviation, meaning all users answered every question. 
% In contrast, CodeQL \emph{Task 1} had much more variation ($std. dev. = 1.73$).
% \by{Text update about statistical tests}
% \hj{TODO; stick to the submitted paper's version of the text, but report the p-values and stdev}

\begin{shadedbox}
\noindent
%Participants provided more correct answers using \tool{} than the baseline. When needing to reason multiple taint flows, their answers were 50\% more accurate.
Participants gave more correct and complete answers using \tool{} than the baseline. Their answers were over 50\% more accurate and significantly better ($p < 0.05$). \tool{} users also left fewer questions unanswered.
%\by{Updated the summary box}
\end{shadedbox}

\subsection{RQ2. Cognitive Load and Confidence} 
% \begin{table}[h!]
% \centering
% \begin{tabular}{l|cc}
% \hline
% \textbf{Dimension} & \textbf{CodeQL (avg)} & \textbf{TraceLens (avg)} \\
% \hline
% Mental Demand       & 5.92                  & 3.25                   \\
% Hurriedness         & 6.17                  & 2.83                   \\
% Perceived Success   & 3.08                  & 6.17                   \\
% Effort              & 6.08                  & 2.92                   \\
% Stress              & 5.17                  & 2.33                   \\
% \hline
% \end{tabular}
% \caption{NASA-TLX style ratings for CodeQL vs.\ TraceLens. Lower is more favorable on all dimensions except perceived success (where higher is better).}
% \label{tab:TraceLens_nasatlx}
% \end{table}

% \subsubsection{Comparison of Workload Dimensions}
% \hj{is "workload dimensions" the right word? isn't it the TLX scores about "cognitive load"?} \by{I think it makes sense? Cognitive load is just one sub-metric in the TLX metrics}
% \hj{you're right! I confused myself}
After each task, participants filled in the NASA-TLX questionnaire, which measures five dimensions: mental demand, hurriedness, perceived success, effort, and stress. Figure~\ref{fig:nasa-tlx} reports the average scores for the participants. 
Other than perceived success (where higher is better), lower values are better. All of the NASA-TLX dimensions are rated on a scale from 1 to 7.
Overall, participants found completing the task with \tool{} less mentally demanding (3.25 compared to 5.92), less hurried (2.83 compared to 6.17), and less stressful (2.33 compared to 5.17), while reporting higher perceived success (6.17 compared to 3.08) and requiring less effort (2.92 compared to 6.08). 
% \textcolor{red}{Using \tool{} improves over CodeQL on all five NASA-TLX dimensions.} 
On all NASA-TLX dimensions, the improvements were statistically significant  ($p < 0.05$) with a large effect size (Cohen's $d > 2$ for all five dimensions).
We therefore reject the null hypothesis for RQ2: \tool{} significantly reduces cognitive load and improves perceived success compared to CodeQL.
% Standard deviations for all categories were under $1.19$. 
% and \emph{paired t-tests}. 
% Notably, \emph{Perceived Success} showed the strongest effect (t-test $p = 2.5e-10$), followed by \emph{Hurriedness} ($p = 2.5e-07$) and \emph{Effort} ($p = 1.8e-07$). 
% Standard deviations were moderate across categories: Hurriedness and Stress showed the most variability (both $std. dev. \approx 1.19$), while \emph{Perceived Success} was relatively consistent ($std. dev. \approx 0.67$). 
% These results highlight that TraceLens not only reduced mental and emotional burden, but also delivered a consistently better user experience.
%\by{Text update about statistical tests}

Several participants commented that CodeQL visualizer forced them to “manually inspect the nodes in each path” (P2) and “go back and forth to visualize in [their] head” (P10). 
In contrast, P5 described {\tool}’s approach as “very clear,” 
and P11 noted it was “easier to track the dataflow.” 
These comments reinforce our findings that {\tool}’s inquiry-based debugging reduced the cognitive overhead of analyzing multiple  interconnected flows.
Most participants preferred {\tool}’s graph-based interface over the default list-view of CodeQL visualizer.
The participants identified areas for improvement for \tool{}. 
P9 mentioned that the reliance on node IDs might be unrealistic in real-world scenarios where users do not have those IDs at hand. 
P1, while feeling “more confident using [\tool{}],” pointed out that some “general taint analysis questions” might still be faster in CodeQL visualizer’s tabular format. 
P7 suggested “showing fully qualified names when hovering on a node” to reduce confusion in large, complex graphs.

Participants valued the ease of tracing flows, especially when there are multiple source-sink relationships. 
They envisioned advanced filtering options, improved naming (to avoid codebase ambiguities), and possible integration of textual or tabular summaries.

\begin{shadedbox}
\noindent
%Using {\tool}, participants reported lower cognitive mental demand and found visually clear graphs and template queries useful during their sensemaking process.
Using {\tool}, participants reported significantly lower mental demand, effort, stress, and hurriedness, and felt more successful. They found the visually clear graphs and template queries helpful in supporting a smoother and more confident sensemaking process.
%\by{Updated the summary box}
\end{shadedbox}

\subsection{RQ3. Usability and Workflow} 

\subsubsection{Interface Features}
The usefulness of {\tool}’s interface was rated by the participants, beyond its core queries. 
%Table~\ref{tab:ui_ratings} summarizes these scores (1--5 on 5-point Likert scale). 
Participants rated color-coding with the highest average score (4.83/5).
% , though a few participants felt red and orange could be more distinct in complex diagrams (ID~2). 
P9 said, 
\emph{``Imagine not having the coloring and having to click each node to figure out what they are... The visualization hides the textual noise.''}

Some participants underutilized the expandable taint paths (with an average score of 3.33/5) as they were unaware of the feature.
Meanwhile, P7 wanted explicit labeling of fully qualified names on the graph to avoid switching to the IDE.
Participants found clicking on nodes to jump into code less useful (with an average score of 3.92/5), mainly using it to confirm ambiguous method names.

\begin{table}
\centering
\resizebox{\columnwidth}{!}{%
\begin{tabular}{l|c|c|c|c}
\hline
 & \multicolumn{2}{c|}{\textbf{\tool{}}}  & \multicolumn{2}{c}{\textbf{CodeQL visualizer}} \\
\textbf{Q} & \textbf{Task 1 (C/E)} & \textbf{Task 2 (C/E)} & \textbf{Task 1 (C/E)} & \textbf{Task 2 (C/E)} \\
\hline
(1) & 100\%/0\% & 100\%/0\% & 100\%/0\% & 67\%/0\% \\
(2) & 83\%/0\%  & 100\%/0\% & 83\%/0\%  & 50\%/0\% \\
(3) & 33\%/0\%  & 50\%/0\%  & 17\%/17\% & 0\%/17\% \\
(4) & 100\%/0\% & 100\%/0\% & 83\%/0\%  & 83\%/17\% \\
(5) & 67\%/0\%  & 50\%/0\%  & 50\%/50\% & 83\%/17\% \\
(6) & 67\%/0\%  & 83\%/0\%  & 17\%/67\% & 33\%/17\% \\
(7) & 100\%/0\% & 67\%/17\% & 83\%/17\% & 17\%/83\% \\
(8) & 100\%/0\% & 67\%/17\% & 67\%/33\% & 17\%/83\% \\
\hline
\end{tabular}%
}
\caption{User Study Results: Percentage of participants providing (C)orrect/(E)mpty answers when using \tool{} vs. CodeQL visualizer.}
\label{tab:tracelens_accuracy}
\end{table}

\subsubsection{Confidence and Ease of Use}

Table~\ref{tab:confidence_ease} shows participants' ratings of \tool{} and CodeQL visualizer in terms of confidence (in the answers to the questions) and overall ease of use.
\tool{} scored 4.25/5 on both measures, substantially higher than CodeQL visualizer (2.08/5 for confidence, 1.92/5 for ease).
These improvements were statistically significant ($p < 0.05$, Cohen's $d > 2$ for both measures).
We therefore reject the null hypothesis for RQ3: \tool{} significantly improves user confidence and ease of use.
% compared to CodeQL.
% \hj{TODO: both confidence and ease-of-use are significant. update the text}
% \by{Updated.}

Multiple participants appreciated \tool{}’s \emph{``dedicated queries''} (P2) for analyzing the results, noting that \emph{“it makes it easier to handle higher-level tasks”} (P4). 
However, some participants (P1, P9) commented that CodeQL visualizer’s table-based listings would still be useful in simpler scenarios.

\begin{table}[h!]
\centering
\begin{tabular}{l|cc}
\hline
 & \textbf{\tool{}} & \textbf{CodeQL Visualizer} \\
\hline
Confidence (1--5) & 4.25 & 2.08 \\
Ease of Use (1--5) & 4.25 & 1.92 \\
\hline
\end{tabular}
\caption{The participants were confident in answering questions with \tool{} and found it easy to use.}
\label{tab:confidence_ease}
\end{table}

% \subsubsection{Suggested Improvements}

% While participants generally found TraceLens “less cumbersome” (P2) for analysis involving multiple flows, they provided suggestions for improving \tool{}:
% \begin{itemize}
%     \item Adding a legend or overlay for color-coding and dotted edges (P4, P13).
%     % \item Using different shapes for source vs.\ sink nodes (P2).
%     \item Providing a hybrid text-based or tabular interface alongside the graph (P1, P2, P9).
%     % \item Offering multi-tab or multi-window usage for comparing flows (P6).
%     \item Enabling ``click-to-edit'' sanitizers, which would update downstream flows in real time (P7).
%     \item Integrating TraceLens into an IDE, such that right-clicking on a method reveals possible sinks or sanitizers (P9).
    % \item Improving GlobalImpact’s frequency estimates (P9), for instance by inferring real-world call likelihood.
% \end{itemize}

% Some participants (P1, P10) proposed highlighting lines in the IDE with the same colors shown in TraceLens’s graph, making transitions between code and visualization more seamless.

\subsubsection{Future Adoption}

Participants gave an average rating of 4.83/5, when asked if they would like to use \tool{} in future taint-flow inspections. Many emphasized that despite needing refinements, the specialized queries, graph-based results, and color-coded visualization saved significant time.
P5 commented, \emph{``For a large program, it’s difficult for the programmer to check each path by hand—\tool{}’s approach can save time and reduce errors.''}

\begin{shadedbox}
\noindent
Participants valued {\tool}'s color-coding and graph-based sensemaking features.
% However, they also suggested integrating textual or tabular details, improving node labels, and providing interactive features (e.g., "click-to-edit" sanitizers).
% Overall, the data indicates strong interest in adopting TraceLens's features for larger-scale static analysis scenarios.
\end{shadedbox}

\subsection{Follow-up Study}
\label{casestudy}
To validate our findings from an industry perspective, we conducted studies with two \textbf{security} professionals with 5/5 familiarity in static taint analysis and 7-10 years of experience, using the protocol from Section~\ref{sec:study-protocol}. Both provided more correct answers with \tool{} than CodeQL and preferred \tool{}'s workflow. P101 noted, \textit{CodeQL requires lots of manual labor, whereas \tool{}'s visual components make it easier to navigate. [The workflow] was very intuitive, the tooling eases manual inspection load completely.''} P102 added, \textit{With CodeQL, I wonder whether I'm missing another row that describes the same location.''} Both achieved high task success with both tools (6/7) but reported higher confidence and ease of use with \tool{}. P102 experienced equally low mental demand but felt more rushed with \tool{} (4/7 vs. 2/7), despite lower effort (2/7 vs. 3/7) and no additional frustration. P101 reported reduced mental demand (2/7 vs. 7/7), hurriedness (1/7 vs. 7/7), and effort (2/7 vs. 5/7) with \tool{}, plus slightly less frustration. Overall, \tool{} matched CodeQL's perceived success while reducing cognitive burden.

\section{Discussion}\label{sec:discussion}

\begin{listing}[t]
\caption{   
A \emph{divergent path} query flagging \emph{branch points} where flow diverges to multiple targets. 
%Using the raw three‐arity \texttt{edge(id,src,dst)} input, we declare \texttt{branch(n)} for each $n$ that appears as \texttt{src} in more than one \texttt{edge(\_,n,\_)} fact.
}
\label{lst:query-extension}
\centering
\begin{lstlisting}[
  language=Prolog,         % Closest match to cplint
  numbers=left,            % Line numbers
  breaklines=true,         % Wrap long lines
  breakatwhitespace=false, % Allow breaking anywhere
  basicstyle=\ttfamily\small,
  xleftmargin=0pt,
  frame=lines
]
.decl edge(id: number, src: number, dst: number)
.input edge

.decl branch(n: number)
.output branch
branch(n) :-
    edge(_, n, _),             
    c = count : edge(_, n, _), 
    c > 1.                     
\end{lstlisting}
\end{listing}

\textit{Improved Accuracy with Visual Queries.}
Using \tool{}, participants achieved better accuracy, especially when analyzing multiple flows. In particular, the AffectedSinks and GlobalImpact queries require the analysis of multiple  flows. 
Surprisingly, participants were able to answer questions related to ``why-not'' using both tools.
% to find all unreachable sinks in AffectedSinks and count all API usages for GlobalImpact.
Participants using the CodeQL visualizer frequently had empty submissions due to lack of time.
% , suggesting viewing results one by one in a list view is time-consuming.
% Participants indicated that they would like to use \tool{} features for inspecting taint analysis in the future.
This shows that an interactive, inquiry-based debugger 
% of taint analysis warnings 
has the potential to improve sensemaking for large result sets, which is known to hinder adoption~\cite{johnson2013don}.  

\paragraph{Suggestions for Enhancement.}
While participants found \tool{} ``less cumbersome'' (P2) for analysis involving multiple flows, they offered suggestions for improving \tool{}.
They recommended adding an on-screen legend to clarify colors and dotted edges to avoid visual confusion.
% , and using distinct shapes for different node types (e.g., sources versus sinks) to avoid visual confusion. 
Additional interactive features, such as ``click-to-edit'' functionality for enabling or disabling sanitizers were also proposed. This suggests that participants would appreciate real-time feedback through interaction on the graph.

\paragraph{Extensibility of Template Queries.}
While \tool{} supports six template questions from Table \ref{tab:queries}, adding more template queries is easy since \tool{}'s template questions are Datalog-style queries.
As a case study, adding a new template query called `divergent path' query to \tool{} (Listing~\ref{lst:query-extension}) took under ten minutes.
This query identifies the \emph{branch points}, which are locations where the flow branches into multiple targets.
Note that \tool{}'s queries support any taint analysis results from CodeQL. 
% More extension queries are available in our replication package (\url{https://github.com/whyflowtaint/WhyFlow/tree/main/data/extension_queries}). 

\paragraph{Supporting Interactivity.}
\tool{} supports interrogative and speculative debugging built on a rich history of using logic programming for software comprehension~\cite{holt_1998, mens_2002, hajiyev_2005, eichberg_2008, guéhéneuc_2008, tourwe_2003}. 
% Soufflé is the state of the art logic query engine.
\tool{} does not compete with underlying taint analysis engines but instead emphasizes~\emph{sensemaking} support.

\section{Threats to Validity}\label{sec:threats}
\paragraph{Construct Validity.}
We designed questions based on realistic ``why,'' ``why-not,'' and ``what-if'' debugging scenarios~\cite{ko2004designing}, requiring reasoning about multiple flows rather than simple lookup tasks.
% We measured cognitive load using the validated NASA-TLX questionnaire~\cite{hart1988development} and assessed confidence and ease of use based on the technology acceptance model~\cite{lee2003technology}.
We employed the NASA-TLX questionnaire~\cite{hart1988development} and the technology acceptance model~\cite{lee2003technology}.
However, 
% the multiple-choice format may not fully capture open-ended debugging, and 
the self-reported metrics may be influenced by subjective interpretations.
\paragraph{Internal Validity.}
We employed a within-subject crossover design with counterbalanced tool order and task assignment to mitigate learning effects and order bias~\cite{vegas2015crossover}.
However, participants may experience fatigue, and time pressure. 
% (20 minutes per task may disadvantage CodeQL's list-based interface).
% , or tutorial effectiveness differences.
The counterbalanced assignment controls for difficulty differences between problem sets, but residual variations could affect results.
\paragraph{External Validity.}
Although our study included only 12 participants, this aligns with typical sample sizes in within-subject SE and HCI studies~\cite{vegas2015crossover, garcia_2024, kang_2024, ganji_2023, horvath_2022, suh_2023, huh_2024}, which require fewer participants than between-subject designs. Most participants were students, but prior work shows that students perform comparably to professionals on security tasks~\cite{sandoval_2023, acar_2016, acar_2017, salman_2015, naiakshina_2017, naiakshina_2020, ko_2015}, and a follow-up study with two security professionals (Section~\ref{casestudy}) further corroborated our findings.

While our study evaluated \tool{} only against CodeQL on Apache Dubbo, 
%our approach generalizes to any taint analyzer: 
\tool{} operates on generic dataflow facts (nodes, edges, sources, sinks) extractable from any analyzer, template queries are Datalog rules independent of the analysis engine, and the approach is extensible without core modifications (Section~\ref{sec:discussion}).
% However, evaluation on additional subject programs and analyzers would strengthen generalizability.

% As stated in Section~\ref{sec:approach}, \tool{} assumes users can recognize suspicious flows, operates atop unmodified CodeQL results with third-party library models, and focuses on end-user debugging rather than false positive/negative classification.
% These assumptions align with practical taint-analysis configuration maintenance.
% but may limit applicability where models are unavailable or users lack domain knowledge.

\section{Related Work}\label{sec:relatedwork}
\textbf{Speculative What-If Analysis.}
Speculative execution~\cite{brun2010speculative,brun2011proactive} allows the investigation of future or alternative actions developers may perform, such as for 
% Brun et al. explored its application for 
providing fix suggestions~\cite{brun2010speculative} and preventing merge conflicts~\cite{brun2011proactive}.
% \sout{, as well as identifying conflicts during version merging~\cite{brun2011proactive}}.
% Prior work applied speculative analysis on version merging, program repair.
Our work is the first to apply it for end-user debugging of taint analysis.
 % for reasoning about permissible dataflows of sensitive information under different configurations of dataflow connectivity.

\textbf{Templated Questions.}
Developers often ask questions during a variety of developer tasks~\cite{begel2014analyze}.
In particular, developers need support for understanding code reachability~\cite{ko2004designing}, exploring how different vulnerabilities relate and finding similar ones~\cite{smith2015questions}, tracking intermediate states of analysis~\cite{do2018debugging}, reasoning about system-wide implications when making changes~\cite{smith2015questions}.
\tool{} is the first work providing templated questions for investigating and debugging the results of taint analysis.

\textbf{Modelling Third-Party Libraries.}
Existing work~\cite{li2024llm,piskachev2019swan_assist, Chiang_Li_Zhou_Banerjee_Schaef_Lyu_Nguyen_Tripp_2024} automatically infers models of third-party libraries.
However, they do not guide users to understand the impact of these models.
% \sout{.
% They do not support end-user debugging and do not allow users to ask questions about the impact of modelling choices}.
While Paralib~\cite{yan2022concept} allows comparison between multiple library choices, it does not allow reasoning about how incorrect models cause a mismatch between user expectations and the actual tool results.

\textbf{End-user debugging.}
End user debugging~\cite{burnett2004end,kissinger2006supporting,grigoreanu2012end} focuses on investigating the root causes of tool outcomes, and has been applied to spreadsheet debugging, etc. 
\tool{} is an end-user debugger for reasoning how taint flows are affected by the configuration of sources, sinks, and models of third-party libraries.

\section{Conclusion}\label{sec:conclusion}
We presented \tool{}, a tool for end-user debugging for taint analysis. 
Through speculative analysis,
\tool{} allows developers to ask ``why'', ``why-not'', and ``what-if'' questions about the analysis configuration, and is able to visualize multiple, interconnected flows on a graphical view.
% designed to mitigate inaccuracies stemming from over- or under-approximate taint specifications in third-party libraries. 
% By enabling developers to pose targeted ``why'' and ``why-not'' questions about data flows, 
\tool{} enables sensemaking of taint analysis and helps users identify root causes of unexpected flows or missing flows. 
% Finally, when developers mark specific flows as correct or incorrect, \tool{} can automatically pinpoint which library specifications must be adjusted to rectify the analysis.
Our user study confirms that \tool{}
helps users provide 21\% more correct answers to questions about taint analysis, while experiencing a lower cognitive load. 
% significantly enhances the debugging experience compared to a baseline interface. 
% Across multiple case studies, participants who used \tool{} were \textbf{30\%} more likely to complete their assigned tasks and up to \textbf{3 times} more likely to find and explain the underlying causes of incorrect flows. 
These results suggest that \tool{}'s inquiry-based approach empowers developers to analyze the outputs of taint analysis.

\section*{Data Availability}
The replication package and study data are available at \url{https://github.com/UCLA-SEAL/WhyFlow}.

\begin{acks}
This work is supported by the National Science Foundation under grant numbers 2426162, 2106838, and 2106404. 
It is also supported in part by funding from Amazon and Samsung. 
We thank the anonymous reviewers for their constructive feedback that helped improve the work.
\end{acks}

% \balance
\bibliographystyle{IEEEtran}
\bibliography{main}

\end{document}